\documentclass[twocolumn,showpacs,aps,floatfix,prd,nofootinbib]{revtex4}
\usepackage{graphicx}
\usepackage{dcolumn}
\usepackage{array, longtable}
\usepackage{epsfig}
\usepackage{amsmath}
\usepackage{colordvi}
\usepackage{hhline}

\newcommand{\BaBarYear}{2005}
\newcommand{\BaBarNumber}{1164}
\newcommand{\SLACPubNumber}{11663}

 \newcommand{\BaBarType}      {PUB}  

\input pubboard/babarsym

\long\def\inst#1{\par\nobreak\kern 4pt\nobreak
    {\it #1}\par\vskip 10pt plus 3pt minus 3pt}

\begin{document}

%


\begin{flushleft}
\babar-\BaBarType-\BaBarYear/\BaBarNumber \\
SLAC-PUB-\SLACPubNumber
\end{flushleft}


\title{\large \bf
\boldmath
The $\epem\to 3(\pipi)$, $2(\pipi\pi^0)$  and                 
$K^+K^- 2(\pipi)$
Cross Sections at Center-of-Mass Energies 
from Production Threshold to
4.5~\gev Measured with
Initial-State Radiation 
} 

%
\author{B.~Aubert}
\author{R.~Barate}
\author{D.~Boutigny}
\author{F.~Couderc}
\author{Y.~Karyotakis}
\author{J.~P.~Lees}
\author{V.~Poireau}
\author{V.~Tisserand}
\author{A.~Zghiche}
\affiliation{Laboratoire de Physique des Particules, F-74941 Annecy-le-Vieux, France }
\author{E.~Grauges}
\affiliation{IFAE, Universitat Autonoma de Barcelona, E-08193 Bellaterra, Barcelona, Spain }
% ...
\collaboration{The \babar\ Collaboration}
\noaffiliation

\date{\today}

\begin{abstract}
We study the processes $\epem\to 3(\pipi)\gamma$,
$2(\pipi\pi^0)\gamma$ and  $K^+ K^- 2(\pipi)\gamma$,  with the 
photon radiated from the initial state.  About 20,000, 33,000 and 4,000 fully
reconstructed events, respectively, have been selected from 232~\invfb of \babar\
data. The
invariant mass of the hadronic final state defines the effective \epem
center-of-mass energy, so that these
data can be compared with 
the corresponding direct \epem measurements. From the $3(\pipi)$,
$2(\pipi\pi^0)$ and $K^+ K^- 2(\pipi)$ mass spectra, the
cross sections for the processes $\epem\to 3(\pipi)$, $\epem\to
2(\pipi\pi^0)$ and $\epem\to K^+ K^- 2(\pipi)$ are measured
for center-of-mass energies from production threshold to 4.5~\gev. The
uncertainty  in the cross
section measurement is typically 6-15\%. 
We observe the $J/\psi$ in all these final states and measure the
corresponding branching fractions.  
\end{abstract}


\centerline{To be Submitted to Physical Review D}
\maketitle

\setcounter{footnote}{0}

\section{Introduction}
\label{sec:Introduction}

The idea of utilizing initial-state radiation (ISR) from a high-mass state
to explore electron-positron processes at all energies below that state was
outlined in Ref.~\cite{baier}.  The possibility of exploiting such processes
in high luminosity $\phi$- and $B$-factories was discussed in
Refs.~\cite{arbus, kuehn, ivanch} and motivates the
hadronic cross section measurement described in this
paper.  This is of particular interest because of the small deviation
of the measured muon $g-2$ value from that predicted by the Standard
Model~\cite{dehz}, where hadronic loop contributions are obtained from \epem
experiments at low center-of-mass (c.m.\@) energies. The study of ISR events
at $B$-factories provides independent and contiguous measurements of
hadronic cross sections in this energy region and also contributes to the
investigation of low-mass resonance spectroscopy. 
                                
The ISR cross section for a particular hadronic final state $f$ 
is related to the 
corresponding \epem cross section $\sigma_f(s)$ by:
\begin{equation}
\frac{d\sigma_f(s,x)}{dx} = W(s,x)\cdot \sigma_f(s(1-x))\ ,
\label{eq1}
\end{equation}
where $x=2E_{\gamma}/\sqrt{s}$; $E_{\gamma}$ is the
energy of the ISR photon in the nominal \epem c.m.\@ frame; $\sqrt{s}
= E_{c.m.}$ is 
the nominal \epem c.m.\@ energy; and $\sqrt{s(1-x)}$ is the effective c.m.\@ energy
at which the final state $f$ is produced. The
invariant mass of the hadronic final state is used to measure the effective \epem
c.m.\@  energy.
The function $W(s,x)$ is calculated with better than 1\% accuracy 
(see for example Ref.~\cite{ivanch}) and describes the 
probability density function for ISR photon emission.
ISR photons are produced at all angles, with a distribution peaking at
small  angles with respect to the axis of the beams, and are required
to be  detected in the electromagnetic calorimeter (EMC) of the \babar\
detector. The acceptance for such photons is 10--15\%~\cite{ivanch}
depending on applied selections.

An important advantage  of ISR data is that the entire
range of effective c.m.\@ energies is scanned in one experiment. 
This
avoids the relative normalization uncertainties that inevitably arise when
data from different experiments, or from different machine settings, are
combined.
                             
A disadvantage of the ISR measurement is that the mass resolution is
much poorer than can be obtained in direct annihilation.
The resolution and absolute energy scale can be monitored
directly using the measured width and mass of the $J/\psi$ resonance produced in
the reaction $\epem \to J/\psi\gamma$. Using a kinematic fit to this
reaction, we find the resolution to be about 9~\mevcc for 
decays of $J/\psi$ in the $3(\pipi)$ mode and about 15~\mevcc in the
$2(\pipi\pi^0)$ mode as will be shown later.

Studies of $e^+e^-\to\mumu\gamma$  and several multi-hadron ISR processes using
\babar\ data have been reported previously~\cite{Druzhinin1,isr3pi,isr4pi}.
These demonstrated good 
detector efficiency and particle identification capability for events of
this kind.

This paper reports analyses of the $3(\pipi)$, $2(\pipi\pi^0)$ and
$K^+K^- 2(\pipi)$ final states produced in
conjunction with a hard photon, assumed to result from ISR.
A clear $J/\psi$ signal is observed for each of these hadronic final states
and the corresponding $J/\psi$ branching fractions are measured.
While \babar\ data are available at effective c.m.\@ energies up to 10.58 \gev, 
the present analysis is restricted to 
energies below 4.5 \gev because of the increase with energy of the
backgrounds from non-ISR 
multihadron production.

\section{\boldmath The \babar\ detector and dataset}
\label{sec:babar}

The data used in this analysis were collected with the \babar\ detector at
the \pep2\ asymmetric \epem\ storage ring. The total integrated luminosity
used is 232~\invfb, which includes data collected at the $\Upsilon(4S)$
resonance mass (211~\invfb), and at a c.m. energy 40~\mev lower (21~\invfb).

The \babar\ detector is described elsewhere~\cite{babar}. 
Charged particles are reconstructed in the \babar\ tracking system,
which comprises the silicon vertex tracker (SVT) and the drift chamber (DCH).
Separation of pions and kaons is accomplished by means of the detector of
internally reflected cherenkov light (DIRC) and energy-loss measurements in
the SVT and DCH. The hard ISR photon and photons from $\pi^0$ decays are
detected in the electromagnetic calorimeter (EMC).  Muon identification  is provided by the
Instrumented  flux return (IFR).

The initial selection of candidate events requires that a high-energy photon
in the event with $E^\gamma_{\rm c.m.} > 3~\gev$ be found recoiling against
six good-quality charged tracks with zero net charge or against four
good-quality charged tracks with zero net charge and four or more photons with
energy higher than 0.02~\gev. 
Almost every candidate event has extra soft photons with energy above
this threshold, mostly due to secondary hadron interactions and machine background.
Each charged track is
required to originate close to the interaction region, to have transverse
momentum greater than 0.1~\gevc and to have a polar angle in the laboratory
frame with respect to the collision axis in the range from 0.4 to 2.45
radians. These selections guarantee the quality of the charged tracks in the DCH. 
The charged track vertex is used as the point of origin to calculate
the  angles for all detected photons.
Events with electrons and positrons are removed on the basis of associated
EMC energy deposition and energy-loss (\dedx) information from the
DCH. 

In order to study the detector acceptance and efficiency, we developed a
set of simulation programs for radiative processes.
The simulation of the
$3(\pipi )\gamma$ and $2(\pipi\pi^0)\gamma$ final states is based on
the generator developed
according to the approach suggested  
by K\"uhn  and Czy\.z~\cite{kuehn2}.  
For the acceptance study we simulate six-charged pions in a phase
space model and in a model which assumes only one
$\rho(770)$ per event, i.e. a $\rho (770)2(\pipi)$ final state.
The $2(\pipi\pi^0)$ and $K^+ K^- 2(\pipi)$ final states are simulated according to      
phase space. 

Multiple soft-photon emission from the initial-state charged particles is
implemented with the structure-function technique~\cite{kuraev, strfun},
while extra photon radiation from the final-state particles is simulated by
means of the PHOTOS package~\cite{PHOTOS}.  The accuracy of the radiative
corrections is about 1\%.

A sample of about 400,000 events were generated with these tools for each
mode and
passed through the detector response simulation \cite{GEANT4}. These
events were then reconstructed through
the same software chain as the experimental data. Variations in detector
and background conditions were taken into account.

For purposes of background estimation, a large sample of events from the
main ISR processes ($2\pi\gamma$, $3\pi\gamma$ ... $5\pi\gamma$, 
$2K\pi\gamma$ ...) was simulated.  
This sample exceeded the expected number of
events in the dataset by a factor of about three.  
In addition, the 
expected numbers of $\epem\to q \qbar$ $(q = u, d, s, c)$ events  were generated via 
JETSET~\cite{jetset} and $\epem\to\tau^+\tau^-$ via KORALB~\cite{koralb}
in order to estimate 
background contributions from non-ISR events. The cross sections for
the above  processes are known with
about 10\% accuracy or better, which is sufficient for the background  
contribution study.

\section{\boldmath The kinematic fit procedure}
\label{sec:Analysis}
The initial sample of candidate events is subjected to a constrained
kinematic fit in conjunction with charged-particle identification
to extract events corresponding to the final states of
interest.

For each particular six-charged-particle candidate, and for each possible
combination of particle types (i.e. $3(\pipi)$ or $K^+ K^- 2(\pipi)$), a
one-constraint (1C)
kinematic fit is performed without using information from the
detected photon candidate.  The only  constraint used is zero photon mass.
Due to the excellent resolution
of the DCH, the three-momentum vector of the photon is better determined
through momentum conservation than through measurement in the EMC. 
As a consequence, the calibration accuracy of the EMC and its alignment with 
respect to the DCH do not contribute to the systematic uncertainties.  
The initial \epem
and final-state charged-particle four-momenta and their covariance matrices
are taken into account.  

The fit for the six-pion final-state hypothesis is
retained for every event. 
If only one track is identified as a kaon,
or if two oppositely-charged kaons are identified, the
$K^+K^- 2(\pipi)$ fit is also retained. 

For the $2(\pipi\pi^0)\gamma$ events a kinematic
fit is performed using the initial \epem, final-state charged-particle and
photon four-momenta and their covariance matrices. 
The highest c.m. energy photon is assumed to be from ISR. Only the
direction  of the photon momentum vector is
used in the fit, not the measured
energy.  All other photons with energies above 20~\mev are paired.
Combinations  lying within $\pm 35 \mevcc$ of the $\pi^0$ mass are tested, and
the event  combination with the best $\chi^2$ value is retained, subject to
the  additional constraint that the two, two-photon pairs are
consistent with  the $\pi^0$ mass. In total five constraints
(5C fit) are applied.
The three-momentum vectors obtained from
the fit  for each charged track and photon are used in further calculations.

\begin{figure}[t]
\includegraphics[width=0.9\linewidth]{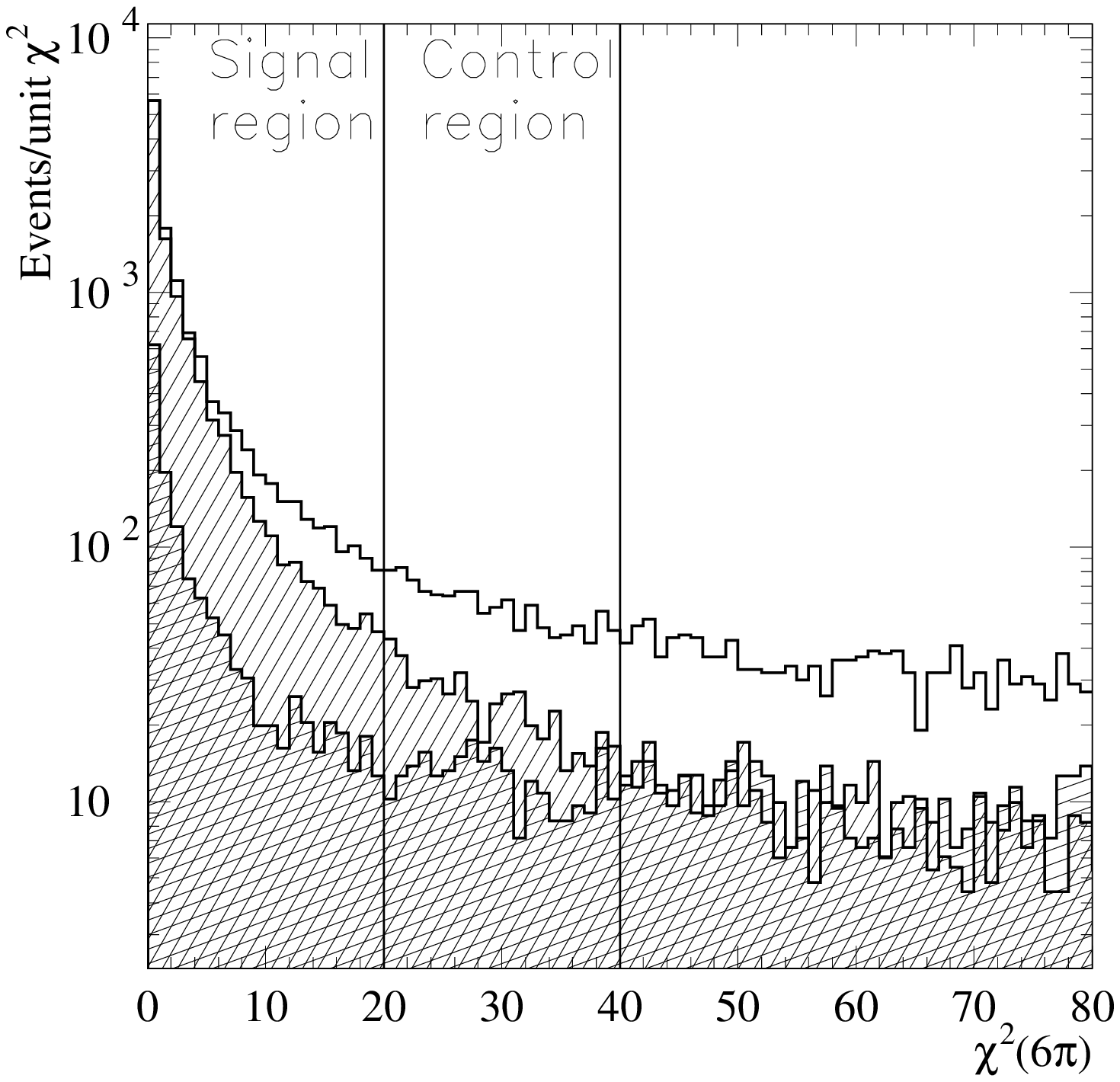}
\vspace{-0.4cm}
\caption{
The one-constraint \chisq distributions for data (unshaded histogram) and
       MC $3(\pipi)\gamma$ simulation (shaded histogram) for six-charged-track events
       fitted to the six-pion hypothesis. The cross-hatched histogram
       is the estimated background contribution from non-ISR events
       obtained from JETSET. The signal and control regions are indicated.
}
\label{6pi_chi2_all}
\end{figure}
\begin{figure}[tbh]
\includegraphics[width=0.9\linewidth]{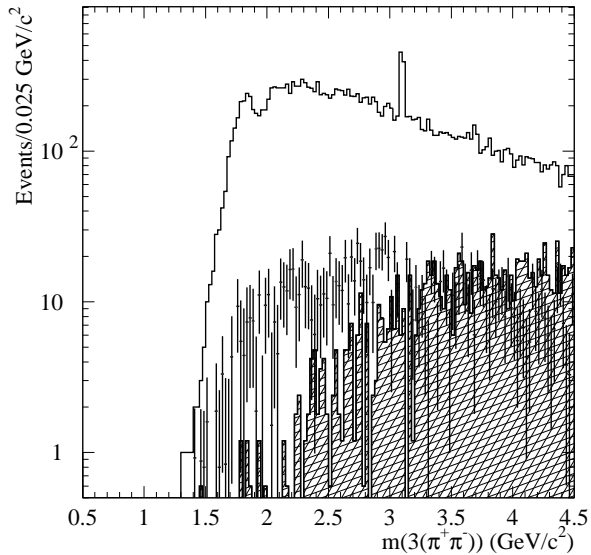}
\vspace{-0.4cm}
\caption{
The six-pion invariant mass distribution (unshaded histogram) for the signal region of 
Fig.~\ref{6pi_chi2_all}. The points indicate the background estimated from
the difference between data and MC events for the control region of
Fig.~\ref{6pi_chi2_all}, normalized to the difference between data and MC
events in
the signal region of Fig.~\ref{6pi_chi2_all}. The cross-hatched histogram
corresponds to the non-ISR background of Fig.~\ref{6pi_chi2_all}.
}
\label{6pi_babar}
\end{figure}

\section{The {\boldmath $3(\pipi)$} final state}
\subsection{Additional selection criteria}

The results of the 1C fit to the six charged-track candidates    
are used to make the final selection of the six-pion sample. 
The momentum vector of the photon reconstructed by
the fit in the laboratory frame is required to have a polar angle $\theta^{\rm
fit}_{\gamma}$ in the range from 0.35 to 2.4~radians and to match the
measured polar angle $\theta^{\rm meas}_{\gamma}$ of the ISR photon in
the EMC within 50 mrad. The corresponding azimuthal angles, $\phi^{\rm
fit}_{\gamma}$ and $\phi^{\rm meas}_{\gamma}$, are also required to agree 
within
this same tolerance. These angular criteria reduce the background by a factor
of about two with no noticeable loss of signal. Finally, the polar angle
$\theta^{\rm fit}_{\rm ch}$ of each charged track obtained from the fit has to
satisfy $0.45<\theta^{\rm fit}_{\rm ch}<2.4$~radians in order to fall within
the acceptance of the DIRC, which provides about 80\% of the kaon 
identification efficiency.

The 1C-fit \chisq distribution for the six-pion candidates is
shown as the upper histogram of Fig.~\ref{6pi_chi2_all}, while the shaded
region is for the corresponding MC-simulated pure $6\pi\gamma$ events.
The experimental distribution has a contribution from background processes,
but the pure $6\pi\gamma$ MC-simulated distribution is also much broader than the 
usual one-constraint \chisq distribution. This is due to multiple soft-photon
emission (detected or not detected)
in the initial state and radiation from the final-state charged
particles, neither of which is taken into account by the constrained fit but
which exist  both in the data and the MC simulation. The MC simulated
\chisq  distribution of Fig.~\ref{6pi_chi2_all}  is normalized
to the data in the region $\chi^2<1$ where
the background contamination and multiple soft-photon emission due to
ISR or FSR is lowest.

The cross-hatched histogram in
Fig.~\ref{6pi_chi2_all} represents the non-ISR
background contribution obtained from the JETSET simulation of
quark-anti-quark production and hadronization and does not exceed 8\%.  

We require $\chi_{6\pi}^2 < 20$ for the six-pion hypothesis, 
and that any accompanying fit to the 
$2K4\pi$ hypothesis have $\chi_{2K4\pi}^2 >20$. 
The subscripts ``$6\pi$'' and ``$2K4\pi$'' here and below refer
to the $3(\pipi)$ and $K^+ K^- 2(\pipi)$ final states exclusively.
We estimate that   these requirements  
reduce the misidentification of $2K4\pi$  events from 11\% to about
2\%, at the  cost of the loss of about 5\% of the signal $6\pi$ events.

The region $20<\chi_{6\pi}^2<40$ is chosen as a control region for the
estimation of background from other ISR and non-ISR multi-hadron reactions. The
procedure followed is described in the next section.

The signal region of Fig.~\ref{6pi_chi2_all} contains 19,683 data             
and 19,980 MC events, while for the control region the corresponding           
numbers are 2,021 and 875 respectively.                                  
\subsection{Background estimation}
\label{sec:background}

The non-ISR background 
contribution to the signal region is obtained from the JETSET MC
simulation,  normalized using the integrated \epem luminosity. The \chisq
distribution for non-ISR events is shown by the cross-hatched histogram
of Fig.~\ref{6pi_chi2_all}. 
The non-ISR background dominates by 
$\epem\to 6 hadrons +\pi^0$ production with photon from $\pi^0$ mistakenly taken as ISR 
photon. 

MC simulation of the $\tau^+\tau^-$ final state and ISR production of
 multi-hadron final states other than $3(\pipi)$ shows 
that such states yield a background in the selected six-pion sample that exhibits a
relatively flat contribution to the $\chi_{6\pi}^2$ distribution.
To validate these estimates of backgrounds with the data, we subtract
 the  MC simulated signal distribution (the shaded histogram of
 Fig.~\ref{6pi_chi2_all}) from  the unshaded one, after
the non-ISR background is subtracted.
The shape of the resulting histogram is well described by
MC simulation of remaining background processes. Its absolute normalisation
is used to estimate the level of those backgrounds in the signal region.

The background contribution to any distribution other
than \chisq is estimated as the difference between the distributions in the
relevant quantity for data and MC events from the control region of
Fig.~\ref{6pi_chi2_all}, normalized to the difference between the number of
data and MC events in the signal region. The non-ISR
background is subtracted separately.

For example, Fig.~\ref{6pi_babar} shows the six-pion invariant mass
distribution up to 4.5~\gevcc for the signal region of
Fig.~\ref{6pi_chi2_all}. The points with error bars show the ISR background
contribution obtained in the manner described from the control region of
Fig.~\ref{6pi_chi2_all}.  The cross-hatched histogram in
Fig.~\ref{6pi_babar} represents the non-ISR
background contribution obtained from the JETSET MC simulation.

Both backgrounds are relatively small at low mass (about 6-8\%), but the non-ISR
background accounts for about 20-25\% of the observed data at approximately
4\gevcc.  

Accounting for uncertainties in cross sections for background processes, 
uncertainties in normalization of events in the control region and 
statistical fluctuations in the number of simulated events,
we estimate that this procedure for background subtraction results
in a systematic uncertainty of less than 3\% in the number of signal
events in the 1.6--3~\gevcc region of six-pion mass, but that it            
increases to 3--5\% in the region above 3~\gevcc.

By selecting a ``background-free'' $6\pi\gamma$ sample with only six charged
tracks and only one photon (about 5\% of events) we can compare \chisq
distributions for data and MC events up to \chisq=1000.  We estimate that
for a $\chi_{6\pi}^2<20$ selection the net signal size should be increased by
$(3\pm2)\%$ to allow for a slight shape difference between the MC and experimental
\chisq distributions.
\begin{figure}[tbh]
\includegraphics[width=0.9\linewidth]{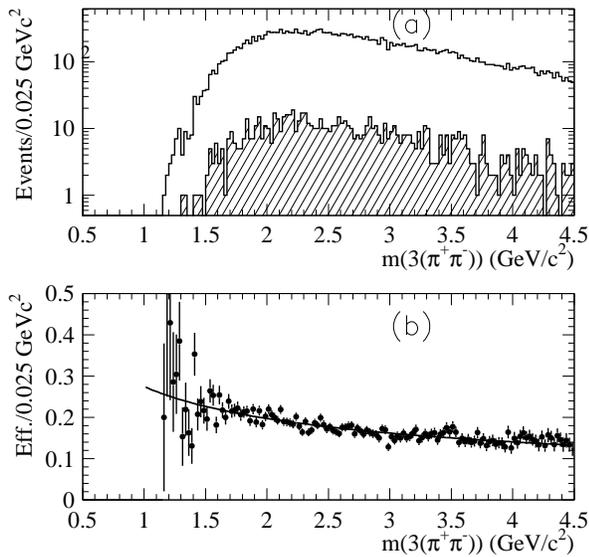}
\vspace{-0.4cm}
\caption{
(a) The six-pion mass distributions from MC simulation for the
  signal (unshaded) and control (shaded) regions of
  Fig.~\ref{6pi_chi2_all}. (b) The mass
  dependence of the net reconstruction and selection efficiency
  obtained from simulation. The curve is a fit described in the text.
}
\label{mc_acc}
\end{figure} 
\subsection{Tracking efficiency}
\label{sec:tracking}
The procedure to measure the track-finding efficiency is described in
our previous paper~\cite{isr4pi} for the four-pion final state. 
The method uses events that
have three charged-particle tracks and a hard
photon. These  events are subjected to a
one-constraint fit, which
uses all measured parameters of the three tracks and the photon and
yields the
three-momentum vector of the missing charged pion in the laboratory frame
assuming this is the only undetected track. If the \chisq of the fit is 
less than 30 and this vector lies within the
acceptance of the DCH, the event is included in the data sample. 
The ratio of three- to four-charged track
events gives the track-finding efficiency. The same procedure is applied to
MC-simulated events. 
The track-finding efficiency is better for MC-simulated events by
$(0.8\pm0.5)\%$ per track independent from momentum~\cite{isr4pi}.
Assuming no increase  in correlations  due to higher 
multiplicity,
we  apply an overall correction of $+(5\pm3)\%$ to the
observed  six-pion event sample based on the previous study.
\subsection{Detection efficiency from simulation}
\label{sec:eff}
\begin{figure}[tbh]
\includegraphics[width=1.1\linewidth]{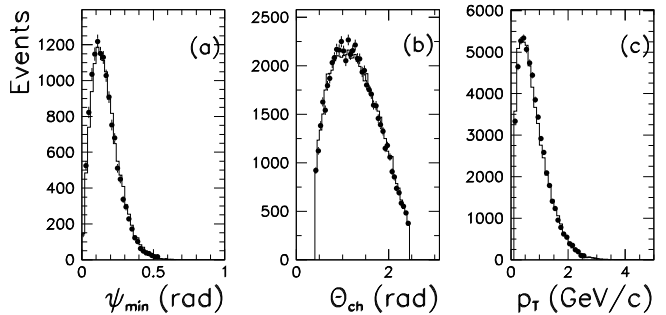}
\vspace{-0.5cm}
\caption{
(a) The distribution in track-pair opening angle for the
minimum of the 15 values possible for each event; (b) the
        distribution in polar angle, and (c) the transverse momentum
        distribution for all pions from all events. All quantities
        are in the laboratory frame; the points are for data and the
        histograms are obtained from MC simulation.
}
\label{mc_distr}
\end{figure}
\begin{figure}[tbh]
\includegraphics[width=1.1\linewidth]{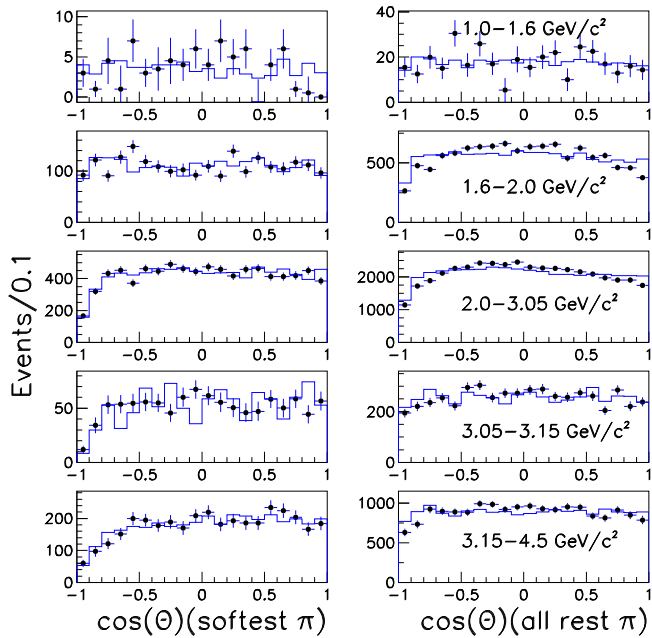}
\vspace{-0.4cm}
\caption{
The angular distribution of the lowest-momentum pion (left)
and of the sum of the remaining five most
energetic pions (right) in the six-pion rest
frame  with respect to the direction of the six-pion system
       in the laboratory frame for the five regions of six-pion mass
       indicated in the right hand plots. The fourth slice is chosen
       to correspond to the $J/\psi$ region.
The points are data, and the histograms are
       MC simulation. 
}
\label{angle_slices}
\end{figure}
The selection procedures applied to the data are also applied to the         
MC-simulated event sample. The resulting six-pion invariant-mass
distribution is shown in Fig.~\ref{mc_acc}(a) for the signal and
control (shaded histogram) regions. 
The mass dependence of the detection
efficiency is obtained by dividing the number of reconstructed MC
events in each 25~\mevcc mass interval by the number generated in          
this same interval. 
Note that the detection 
efficiency computed that way is insensitive to the actual shape of the 
mass distribution of Fig.~\ref{mc_acc}(a) used in MC simulation.
The result is shown in Fig.~\ref{mc_acc}(b); the 
curve is obtained from a 3rd order polynomial fit to the distribution. The              
efficiency falls off gradually 
with increasing mass from about 20\% at 1.6~\gevcc to about 14\% at
4.5~\gevcc.  This efficiency 
estimate takes into account the geometrical acceptance of the detector 
for the final-state photon and the charged pions, the inefficiency of 
the  several detector subsystems, and event-loss due to additional
soft-photon  emission from the initial and final states.

As mentioned in Sec.~\ref{sec:babar}, the model used in the MC simulation 
assumes that the six-pion final state results predominantly from the
$\rho(770) 2(\pipi)$
production process. 
In general, this model describes well the
distributions in many of the kinematic variables characterizing the
six-pion final state. Some examples are shown in Figs.~\ref{mc_distr}
and~\ref{angle_slices}, in which the points with error bars represent
data while the histograms are obtained from MC simulation.
Figure~\ref{mc_distr}(a) shows the distribution in
$\psi_{\rm min}$, the minimum charged-pion-pair opening angle for each
event, while Fig.~\ref{mc_distr}(b) and Fig.~\ref{mc_distr}(c)
represent the distribution in polar angle, $\theta_{\rm ch}$, and
transverse momentum, $p_T$, respectively, for all final-state
pions. All quantities are calculated in the laboratory frame. The        
overall agreement between MC simulation and data is very
good. Figure~\ref{angle_slices} 
compares the distributions in $\cos\theta$, where $\theta$ 
is the angle between a charged pion in the six-pion rest frame, and 
the direction of the six-pion system in the laboratory frame. 
Data and MC are in rather good agreement.

In the six-pion rest frame, the angular acceptance is rather
uniform. 
A simulation without resonances using only
six-pion phase space
does not produce discernible deviations from the observed angular
distributions, and does not change the overall acceptance more than by 3\%. This
value is taken as an estimate of the model-dependent systematic
uncertainty  in the acceptance.

\begin{figure}[tbh]
\includegraphics[width=0.9\linewidth]{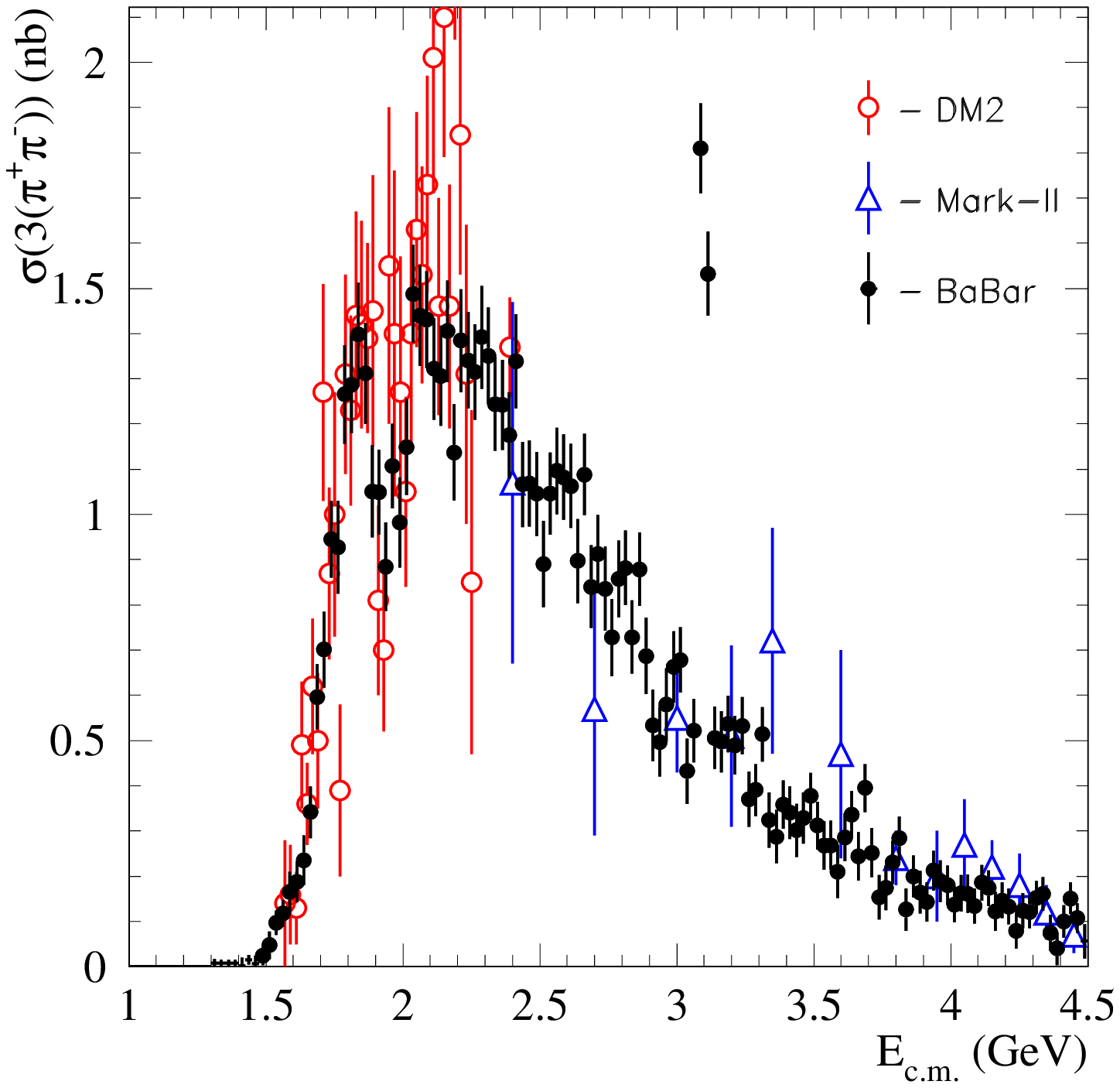}
\vspace{-0.4cm}
\caption{
The \epem c.m.\@ energy dependence of the $3(\pipi )$ cross section
 measured with ISR data at \babar\ compared with the 
direct \epem measurements by DM2 and MARK-II. 
Only statistical errors are shown.
}
\label{6pi_ee_babar}
\end{figure} 
\subsection{\boldmath Cross section for $\epem\to 3(\pipi)$}
Data from the reaction $\epem\to\mumu\gamma$ are used to convert the
invariant-mass distribution for an ISR-produced hadronic final state to        
the energy dependence of the corresponding \epem cross section.
The invariant mass of the muon pair $m_{\rm inv}^{\mu\mu}$ defines an
effective \epem c.m.\@ collision energy, $E_{\rm c.m.}$. The differential
luminosity, $d{\cal L}$, associated with the interval $dE_{\rm c.m.}$ centered at
effective collision energy $E_{\rm c.m.}$ is then obtained from
\begin{equation}
  d{\cal L}(E_{\rm c.m.})
  = \frac{dN_{\mu\mu\gamma}(E_{\rm c.m.})}
         {\epsilon_{\mu\mu}\cdot(1+\delta_{\rm FSR}^{\mu\mu})
          \cdot\sigma_{\mumu}(E_{\rm c.m.})\cdot(1+\delta_{\rm vac})}\  , 
\label{isrlum1}
\end{equation}
where $E_{\rm c.m.}\equiv m_{\rm inv}^{\mu\mu}$; $dN_{\mu\mu\gamma}$ is the number of
muon pairs in the mass interval $dm_{\rm inv}^{\mu\mu}\equiv dE_{\rm c.m.}$; 
$\epsilon_{\mu\mu}$ is the acceptance, corrected for muon identification and
soft-photon emission; $(1+\delta_{\rm FSR}^{\mu\mu})$ corrects for hard photon
emission from final-state muons;
$~\sigma_{\mumu}(E_{\rm c.m.})$  is the $\epem\to\mumu$ Born cross section at 
center-of-mass energy $E_{\rm c.m.}$; and $(1+\delta_{\rm vac})$ is the 
corresponding vacuum
polarization correction~\cite{EJ}.  
The dependence of the differential luminosity on $E_{\rm c.m.}$ is
presented in our previous paper~\cite{isr4pi}. 

From a detailed study of the
$\epem\to\mumu\gamma$ detection and identification efficiency described in
Ref.~\cite{Druzhinin1} and comparison of the observed invariant-mass
spectrum with theoretical calculations, we estimate the systematic uncertainty
associated with luminosity determination to be 3\%.

The six-pion \epem cross section can then be calculated from
\begin{equation}
  \sigma(3(\pipi))(E_{\rm c.m.})
  = \frac{dN_{6\pi\gamma}(E_{\rm c.m.})}
         {d{\cal L}(E_{\rm c.m.})\cdot\epsilon_{6\pi}^{\rm corr}
          \cdot\epsilon_{6\pi}^{\rm MC}(E_{\rm c.m.})}\ ,
\end{equation}
where $m_{\rm inv}^{6\pi} \equiv E_{\rm c.m.}$ with $m_{\rm
  inv}^{6\pi}$ the invariant mass of 
the six-charged-pion system; $dN_{6\pi\gamma}$ is the number of selected
six-charged-pion events after background subtraction in the interval 
$dE_{\rm c.m.}$ and
$\epsilon_{6\pi}^{\rm MC}(E_{\rm c.m.})$ is the corresponding detection
efficiency obtained from the MC simulation. The factor
$\epsilon_{6\pi}^{\rm corr}$ takes into account the 
difference between the \chisq distributions for data and MC events,
and the tracking-efficiency discrepancies 
discussed in Sec.~\ref{sec:background} and Sec.~\ref{sec:tracking} respectively.

\begin{table*}
\caption{Summary of the $\ep\en\to 3(\pipi)$ 
cross section measurement. Errors are statistical only.}
\label{6pi_tab}
\begin{ruledtabular}
\begin{tabular}{ c c c c c c c c }
$E_{\rm c.m.}$ (GeV) & $\sigma$ (nb)  
& $E_{\rm c.m.}$ (GeV) & $\sigma$ (nb) 
& $E_{\rm c.m.}$ (GeV) & $\sigma$ (nb) 
& $E_{\rm c.m.}$ (GeV) & $\sigma$ (nb)  
\\
\hline

 1.3125 &  0.01 $\pm$  0.01 & 2.1125 &  1.36 $\pm$  0.12 & 2.9125 &  0.55 $\pm$  0.08 & 3.7125 &  0.26 $\pm$  0.06 \\
 1.3375 &  0.01 $\pm$  0.01 & 2.1375 &  1.35 $\pm$  0.11 & 2.9375 &  0.51 $\pm$  0.08 & 3.7375 &  0.16 $\pm$  0.05 \\
 1.3625 &  0.01 $\pm$  0.01 & 2.1625 &  1.45 $\pm$  0.12 & 2.9625 &  0.60 $\pm$  0.08 & 3.7625 &  0.18 $\pm$  0.05 \\
 1.3875 &  0.01 $\pm$  0.01 & 2.1875 &  1.17 $\pm$  0.11 & 2.9875 &  0.68 $\pm$  0.08 & 3.7875 &  0.24 $\pm$  0.05 \\
 1.4125 &  0.00 $\pm$  0.02 & 2.2125 &  1.43 $\pm$  0.12 & 3.0125 &  0.70 $\pm$  0.07 & 3.8125 &  0.29 $\pm$  0.05 \\
 1.4375 &  0.02 $\pm$  0.01 & 2.2375 &  1.38 $\pm$  0.11 & 3.0375 &  0.45 $\pm$  0.07 & 3.8375 &  0.13 $\pm$  0.05 \\
 1.4625 &  0.01 $\pm$  0.02 & 2.2625 &  1.36 $\pm$  0.11 & 3.0625 &  0.54 $\pm$  0.07 & 3.8625 &  0.21 $\pm$  0.05 \\
 1.4875 &  0.03 $\pm$  0.02 & 2.2875 &  1.44 $\pm$  0.12 & 3.0875 &  1.87 $\pm$  0.10 & 3.8875 &  0.17 $\pm$  0.05 \\
 1.5125 &  0.05 $\pm$  0.03 & 2.3125 &  1.40 $\pm$  0.11 & 3.1125 &  1.58 $\pm$  0.10 & 3.9125 &  0.15 $\pm$  0.04 \\
 1.5375 &  0.10 $\pm$  0.03 & 2.3375 &  1.28 $\pm$  0.11 & 3.1375 &  0.52 $\pm$  0.07 & 3.9375 &  0.22 $\pm$  0.05 \\
 1.5625 &  0.12 $\pm$  0.03 & 2.3625 &  1.28 $\pm$  0.10 & 3.1625 &  0.51 $\pm$  0.07 & 3.9625 &  0.20 $\pm$  0.05 \\
 1.5875 &  0.17 $\pm$  0.05 & 2.3875 &  1.21 $\pm$  0.10 & 3.1875 &  0.55 $\pm$  0.06 & 3.9875 &  0.19 $\pm$  0.04 \\
 1.6125 &  0.19 $\pm$  0.05 & 2.4125 &  1.38 $\pm$  0.11 & 3.2125 &  0.51 $\pm$  0.07 & 4.0125 &  0.14 $\pm$  0.04 \\
 1.6375 &  0.24 $\pm$  0.06 & 2.4375 &  1.10 $\pm$  0.10 & 3.2375 &  0.55 $\pm$  0.07 & 4.0375 &  0.17 $\pm$  0.04 \\
 1.6625 &  0.35 $\pm$  0.06 & 2.4625 &  1.10 $\pm$  0.10 & 3.2625 &  0.38 $\pm$  0.06 & 4.0625 &  0.17 $\pm$  0.04 \\
 1.6875 &  0.62 $\pm$  0.07 & 2.4875 &  1.08 $\pm$  0.10 & 3.2875 &  0.40 $\pm$  0.06 & 4.0875 &  0.14 $\pm$  0.04 \\
 1.7125 &  0.72 $\pm$  0.09 & 2.5125 &  0.92 $\pm$  0.10 & 3.3125 &  0.53 $\pm$  0.06 & 4.1125 &  0.19 $\pm$  0.04 \\
 1.7375 &  0.98 $\pm$  0.09 & 2.5375 &  1.08 $\pm$  0.09 & 3.3375 &  0.33 $\pm$  0.06 & 4.1375 &  0.18 $\pm$  0.04 \\
 1.7625 &  0.96 $\pm$  0.11 & 2.5625 &  1.13 $\pm$  0.10 & 3.3625 &  0.30 $\pm$  0.06 & 4.1625 &  0.13 $\pm$  0.04 \\
 1.7875 &  1.31 $\pm$  0.11 & 2.5875 &  1.12 $\pm$  0.10 & 3.3875 &  0.37 $\pm$  0.06 & 4.1875 &  0.15 $\pm$  0.04 \\
 1.8125 &  1.33 $\pm$  0.11 & 2.6125 &  1.10 $\pm$  0.10 & 3.4125 &  0.35 $\pm$  0.06 & 4.2125 &  0.14 $\pm$  0.04 \\
 1.8375 &  1.44 $\pm$  0.12 & 2.6375 &  0.93 $\pm$  0.10 & 3.4375 &  0.31 $\pm$  0.06 & 4.2375 &  0.08 $\pm$  0.04 \\
 1.8625 &  1.35 $\pm$  0.12 & 2.6625 &  1.12 $\pm$  0.09 & 3.4625 &  0.34 $\pm$  0.06 & 4.2625 &  0.13 $\pm$  0.04 \\
 1.8875 &  1.09 $\pm$  0.11 & 2.6875 &  0.87 $\pm$  0.09 & 3.4875 &  0.39 $\pm$  0.05 & 4.2875 &  0.13 $\pm$  0.04 \\
 1.9125 &  1.08 $\pm$  0.10 & 2.7125 &  0.94 $\pm$  0.09 & 3.5125 &  0.32 $\pm$  0.05 & 4.3125 &  0.16 $\pm$  0.04 \\
 1.9375 &  0.91 $\pm$  0.10 & 2.7375 &  0.86 $\pm$  0.10 & 3.5375 &  0.28 $\pm$  0.05 & 4.3375 &  0.17 $\pm$  0.04 \\
 1.9625 &  1.14 $\pm$  0.10 & 2.7625 &  0.75 $\pm$  0.09 & 3.5625 &  0.28 $\pm$  0.06 & 4.3625 &  0.08 $\pm$  0.04 \\
 1.9875 &  1.01 $\pm$  0.10 & 2.7875 &  0.89 $\pm$  0.09 & 3.5875 &  0.22 $\pm$  0.06 & 4.3875 &  0.04 $\pm$  0.04 \\
 2.0125 &  1.19 $\pm$  0.11 & 2.8125 &  0.91 $\pm$  0.09 & 3.6125 &  0.30 $\pm$  0.05 & 4.4125 &  0.10 $\pm$  0.04 \\
 2.0375 &  1.54 $\pm$  0.11 & 2.8375 &  0.75 $\pm$  0.08 & 3.6375 &  0.35 $\pm$  0.05 & 4.4375 &  0.16 $\pm$  0.04 \\
 2.0625 &  1.49 $\pm$  0.11 & 2.8625 &  0.91 $\pm$  0.08 & 3.6625 &  0.25 $\pm$  0.05 & 4.4625 &  0.11 $\pm$  0.03 \\
 2.0875 &  1.48 $\pm$  0.11 & 2.8875 &  0.71 $\pm$  0.09 & 3.6875 &  0.41 $\pm$  0.05 & 4.4875 &  0.06 $\pm$  0.04 \\
\end{tabular}
\end{ruledtabular}
\end{table*}

The energy dependence of the cross section for the reaction
$\epem\to 3(\pipi)$ after all corrections is shown in
Fig.~\ref{6pi_ee_babar}. It shows a structure around 1.9~\gev, 
reaches a peak value of about 1.5~nb near 
2.0~\gev, followed by a monotonic decrease toward
higher energies perturbed only by a peak at the $J/\psi$ mass
position.
The cross section for each 25~\mev interval is presented
in Table~\ref{6pi_tab}.  

Since $d{\cal L}$ has been corrected for vacuum polarization 
and final-state soft-photon emission, the six-pion 
cross section measured in this way includes effects due to vacuum 
polarization and final-state soft-photon emission.
For $g-2$ calculations, vacuum polarization
contributions should be excluded from this data. 

We studied the resolution in six-pion mass with MC simulation, 
finding that Gaussian fits of line shapes give mass resolutions $\sigma_{res}$
that vary between 6.2~\mevcc in the 1.5-2.5~\gevcc mass region and 8.7~\mevcc
in the 2.5-3.5~\gevcc mass region. 
The observed line shape is not purely Gaussian mainly due to soft-photon
radiation.  Since the resolution varies slowly with mass, the data are
binned in mass intervals of 25~\mevcc and the cross section has no sharp
peaks (except for the $J/\psi$ region discussed below), unfolding the
resolution has little effect on the measured energy dependence and has not
been performed in this analysis.  For the sake of any future comparisons
of our data with models (e.g. for the structure near 1.9~\gevcc) we
provide the relation between the observed number $N_i$ of events in bin
$i$ and the "true" number of events in nearby bins:
\begin{equation}
N_{i}=e1\cdot N_{i-2}^0 + e2\cdot N_{i-1}^0 + e3\cdot N_{i}^0 + e4\cdot N_{i+1}^0 + 
e5\cdot N_{i+2}^0 ~ ,
\label{masres}
\end{equation}
where $N_{i}^0$ is the number of events in bin $i$ before accounting for
resolution.  The coefficients 
$e1,..e5$=0.005,0.031,0.896,0.062,0.006  
 are the true event
fractions in five energy bins of 25~\mev width for the six-pion mass
region 1.5--2.5~\gevcc.

\subsection{\boldmath Summary of systematic studies}
\label{sec:Systematics}
The measured six-charged-pion cross section values shown in Fig.~\ref{6pi_ee_babar} and 
summarized in Table~\ref{6pi_tab} include only statistical errors. The systematic 
errors discussed in previous sections are summarized in Table~\ref{error_tab}, along 
with the corrections that were applied to the measurements.
\begin{table*}[tbh]
\caption{
Summary of systematic errors for the $\epem\to 3(\pipi)$  cross section measurement.
}
\label{error_tab}
\begin{ruledtabular}
\begin{tabular}{l c l} 
Source & Correction applied & Systematic error\\
\hline
Luminosity from $\mu\mu\gamma$ &  -  &  $3\%$ \\
MC-data difference in $\chi^{2}<20$ signal region & $+3\%$ & $2\%$\\ 
Background subtraction & - &  $3\%$ for $m_{6\pi}<3.0~\gevcc$ \\
                       &   &  $5\%$ for $m_{6\pi}>3.0~\gevcc$ \\
MC-data difference in tracking efficiency & $+5\%$ & $3\%$ \\
Radiative corrections accuracy & - & $1\%$ \\
Acceptance from MC (model-dependent) & - & $3\%$  \\
\hline
Total  (assuming addition in quadrature and no correlations)    &
$+8\%$    & $6\%$ for $m_{6\pi}<3.0~\gevcc$ \\
          &    & $8\%$ for $m_{6\pi}>3.0~\gevcc$\\
\end{tabular}
\end{ruledtabular}
\end{table*}

The two systematic corrections applied to the measured cross sections
sum up to +8\% with 6-8\% taken as a systematic uncertainty.

\subsection{\boldmath Physics results}
\label{sec:Physics}
The six-charged-pion cross section measured by \babar\ can be compared with
existing \epem measurements performed by the DM2~\cite{6pidm2} and 
MARK-II~\cite{6pimark2} detectors (see Fig.~\ref{6pi_ee_babar}). 
The agreement is relatively good, but
the \babar\ measurement is much more precise. The structure around 1.9~\gev reported 
by both the DM2~\cite{6pidm2} and FOCUS~\cite{focus} experiments is clearly seen (see 
discussion later). 

\begin{figure}[tbh]
\includegraphics[width=1.0\linewidth]{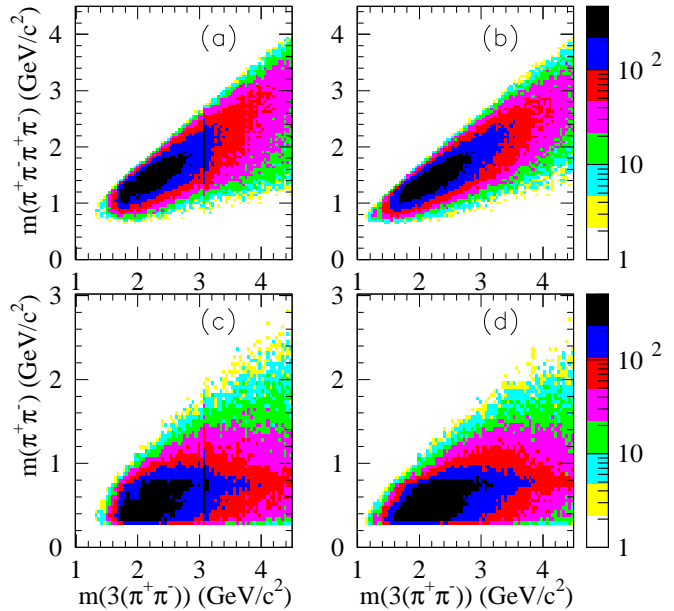}
\vspace{-0.4cm}
\caption{
Invariant mass distributions for all six pion versus neutral four pion
combinations  (top) and neutral two pion combinations (bottom) for
data (a,c)  and MC simulation (b,d). The $J/\psi$ signal seen in the data
is not  included in the simulation.
}
\label{4pi_2-3pi}
\end{figure}

Different mass combinations were studied in data and MC events to       
search for any structures or states not included in the simulation.                
Figure~\ref{4pi_2-3pi} shows the scatter-plots of 4$\pi$- and 2$\pi$-mass
versus 6$\pi$-mass for data and MC events.  Good agreement is seen          
except for narrow regions around the $J/\psi$ and $\psi(2S)$ masses that
are not included in the simulation.                                           

\begin{figure*}[tb]
\vspace{-1.0cm}
\hspace{0.2cm}
\includegraphics[width=0.67\linewidth]{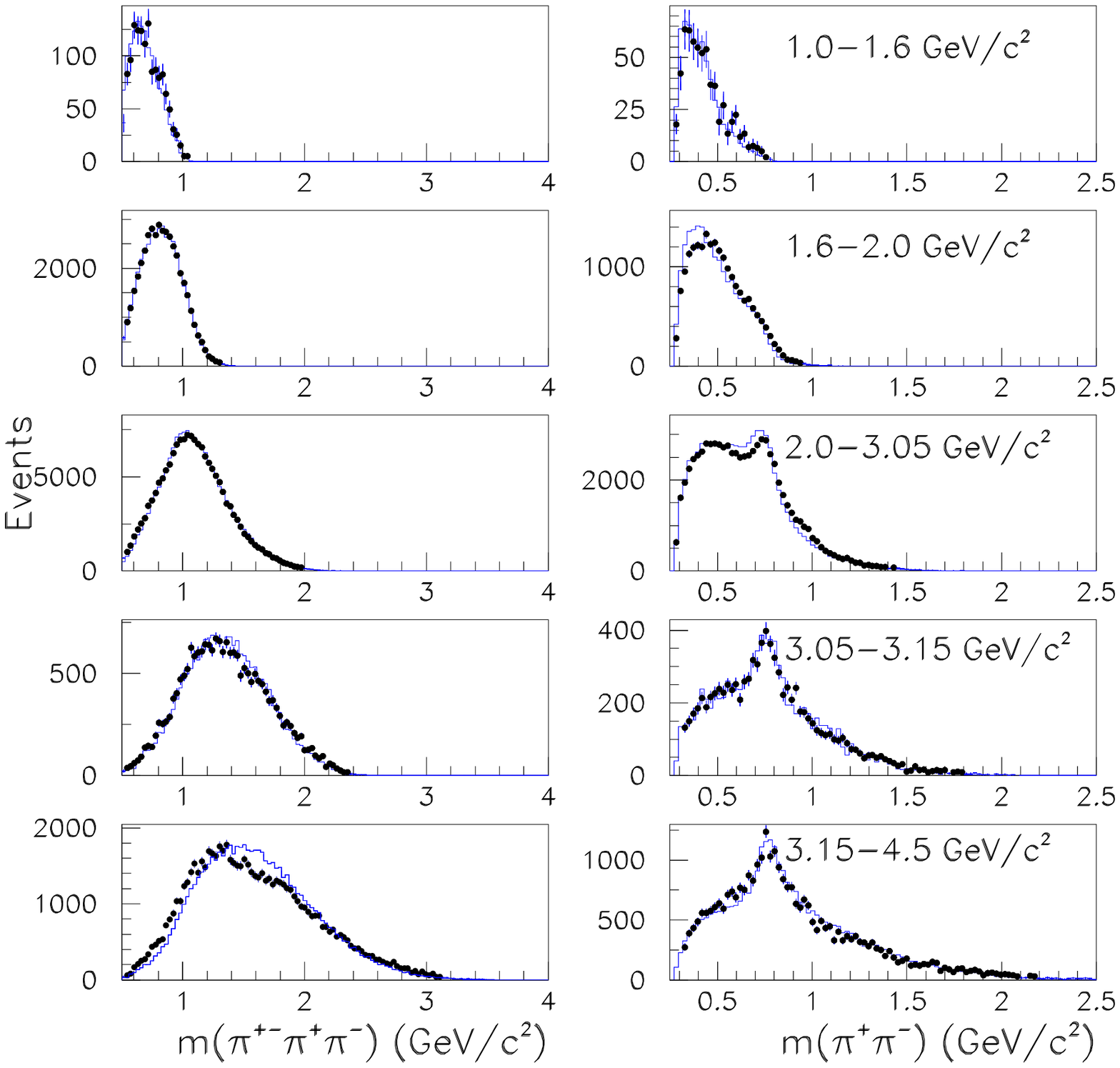}
\hfill
\includegraphics[width=0.307\linewidth]{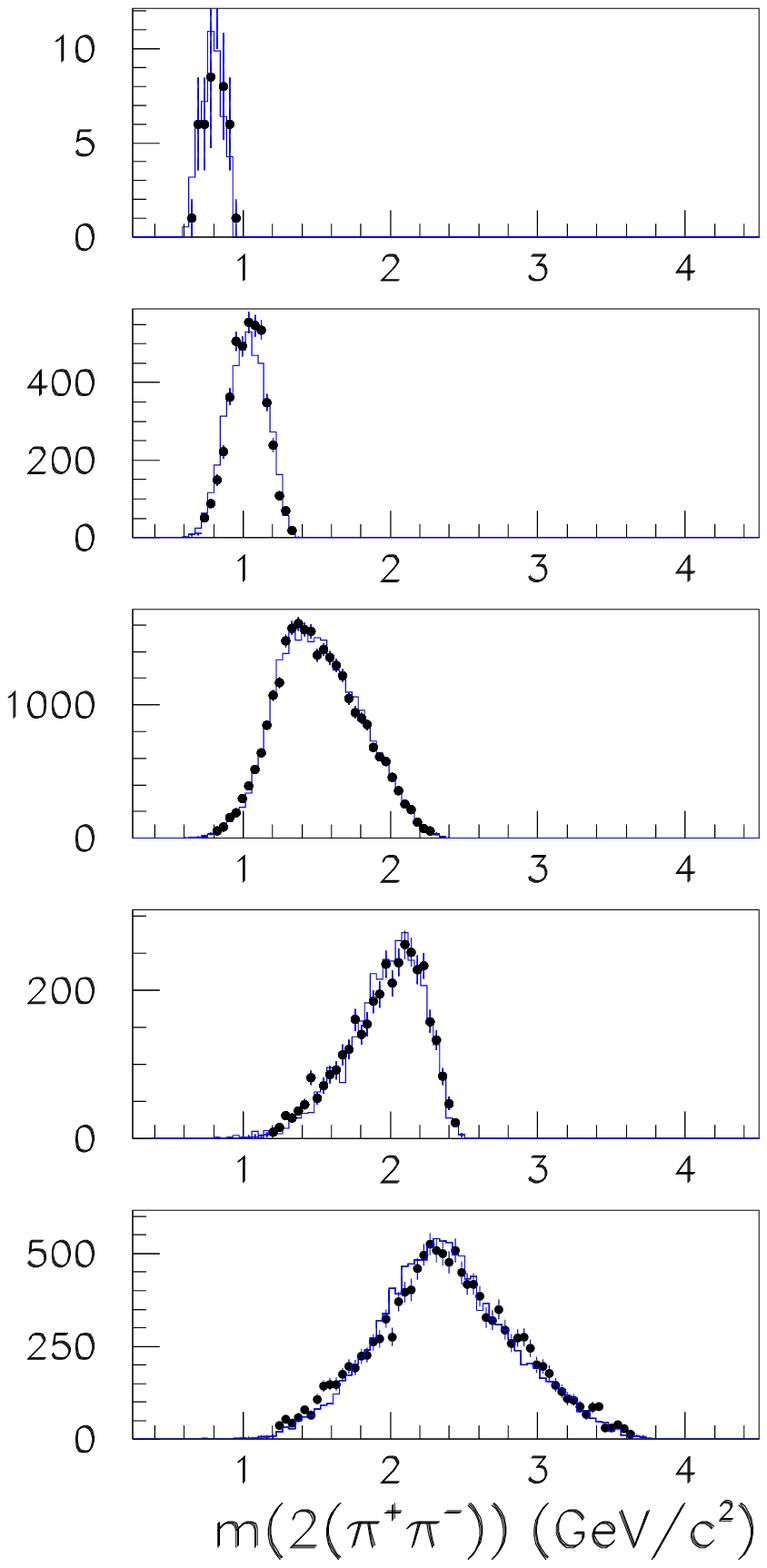}
\vspace{-0.4cm}
\caption{
Invariant mass distributions for different sub-combinations of pions
arranged by  rows for the five different regions of $3(\pipi)$ mass
indicated in  the central column of histograms.  The points
(histograms) display  data (simulation).  The three pion (left column)
and two  pion (middle column) plots sum over all possible
combinations, while  the $2(\pipi)$ combinations (right column) include
only the  remaining combinations from events selected to have one
$\pipi$ in the $\rho(770)$ region.
}
\label{2pi_3pi_pro}
\end{figure*}

In order to make a more detailed study, five intervals of 6$\pi$-mass are
selected: (1) 1.0--1.6~\gevcc; (2) 1.6--2.0~\gevcc
(an interval with significant structure in the cross
section, see Fig.~\ref{6pi_ee_babar}); (3) 2.0--3.05~\gevcc; 
(4) 3.05--3.15~\gevcc (the $J/\psi$ resonance region);
and (5) 3.0--4.5~\gevcc.
Figure~\ref{2pi_3pi_pro}  
shows the two and three pion mass projections, as well as the mass
projections  of the remaining four pions in events having a two pion
combination  within $\pm$75~\mevcc  of the $\rho$ mass, for the five regions of
six-pion mass just described, for both data and
simulation. Background in the  data is subtracted using the \chisq
distributions  as described above.

A simple model with only one $\rho(770)$ per event is in excellent agreement 
with experimental  data. No other significant structures are
observed.
A full partial wave analysis (PWA)  would be required in order to arrive at a more
precise interpretation of the data but the final state with only one $\rho(770)$ 
per event dominates.  This PWA requires a simultaneous
analysis of the $2(\pipi\pi^0)$ final state, which is described
in the next sections.
\begin{figure}[tbh]
\includegraphics[width=0.9\linewidth]{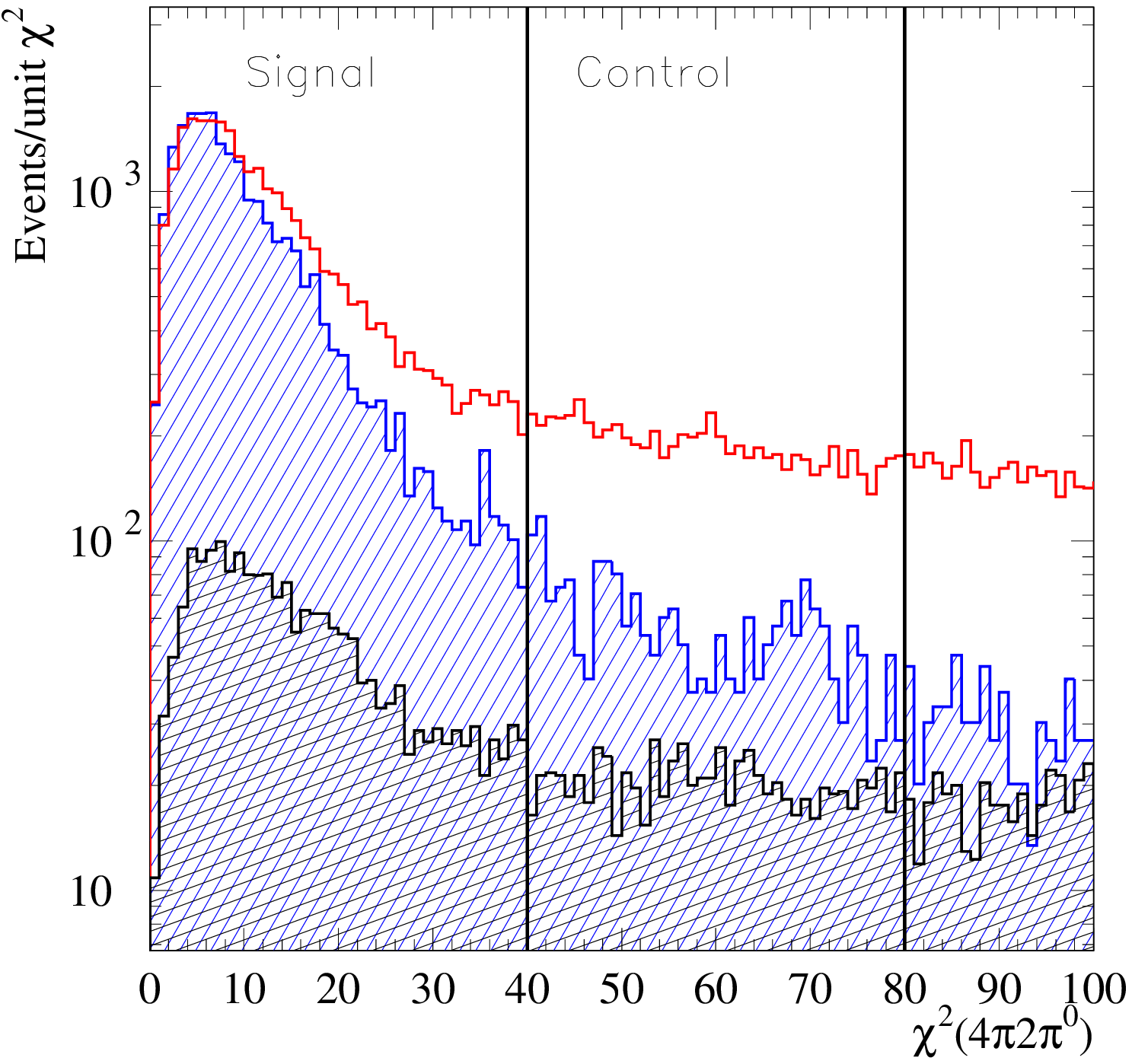}
\vspace{-0.4cm}
\caption{
The five-constraint \chisq distributions for data (unshaded histogram) and
       MC $2(\pipi\pi^0)\gamma$ simulation (shaded) for four-charged-track and
       five-photon events
       fitted to the six-pion hypothesis. The cross-hatched histogram
       is the estimated background contribution from non-ISR events
       obtained from JETSET. The signal and control regions are indicated.
}
\label{4pi2pi0_chi2_all}
\end{figure}
\begin{figure}[tbh]
\includegraphics[width=0.9\linewidth]{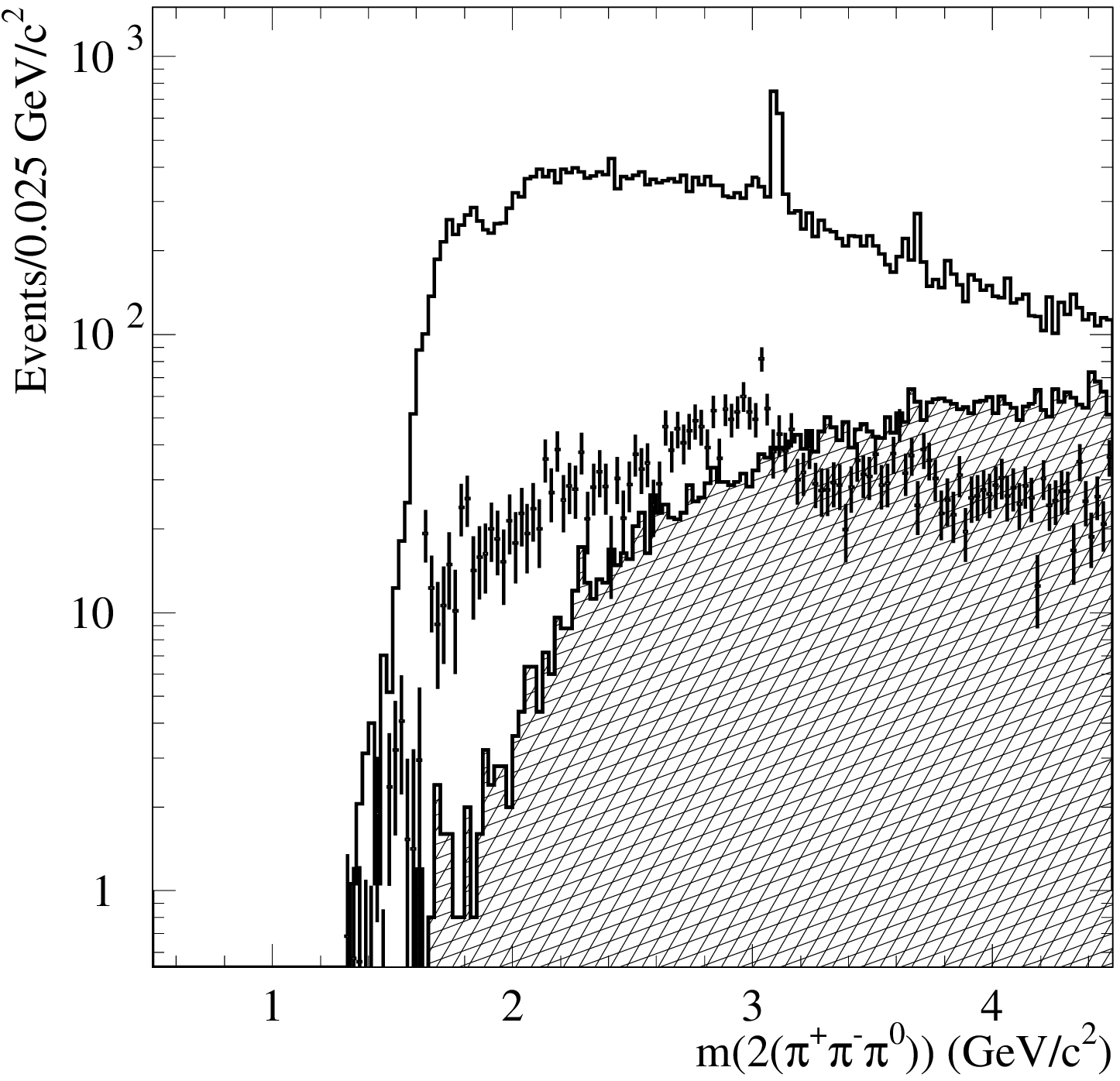}
\vspace{-0.4cm}
\caption{
The six-pion (four charged and two neutral) invariant mass
distribution  (unshaded histogram)
for the signal region of  
Fig.~\ref{4pi2pi0_chi2_all}. The points indicate the background estimated from
the difference between data and MC events for the control region of
Fig.~\ref{4pi2pi0_chi2_all}, normalized to the difference between data and MC
events in
the signal region of Fig.~\ref{4pi2pi0_chi2_all}. The cross-hatched histogram
corresponds to the non-ISR background of Fig.~\ref{4pi2pi0_chi2_all}.
}
\label{4pi2pi0_babar}
\end{figure}
\section{The {\boldmath $2(\pipi\pi^0)$} final state}
\subsection{Additional selection criteria}

The results of the five-constraint fit (see Sec.~\ref{sec:Analysis}) to the
four-charged-track and five-photon candidates
are used to make the final selection of the  $2(\pipi\pi^0)$ sample.
The polar angle
$\theta^{\rm fit}_{\rm ch}$ of each charged track obtained from the fit has to
satisfy $0.45<\theta^{\rm fit}_{\rm ch}<2.4$~radians.
To reduce the background from events where an energetic photon is
coupled with a soft photon from background to form a $\pi^0$, an additional
selection is applied.
The cosine of the helicity angle of the photons of the most energetic $\pi^0$,
measured in the $\pi^0$ rest frame, is required to be less than 0.85.
We further require all photons from $\pi^0$ to have an energy higher than 50~\mev .

After these additional selections we require $\chi_{4\pi 2\pi^0}^2 <
 40$ for the six-pion hypothesis and require 
that none of the charged tracks  be identified as a kaon
to reduce the contribution from $K^+ K^- \pipi\ppz$ and $K K_S 3\pi$ final 
states.

The five-constraint-fit \chisq distribution for the four-pion and
five-photon candidates is
shown as the unshaded histogram of Fig.~\ref{4pi2pi0_chi2_all}, while the shaded
region is for the corresponding MC-simulated pure $2(\pipi\pi^0)\gamma$ events.
The MC-simulated \chisq distribution is
normalized to the data in the region $\chi^2<10$ where contamination of the
background events and multiple soft ISR and FSR is lowest.

The cross-hatched histogram in
Fig.~\ref{4pi2pi0_chi2_all} represents the non-ISR
background contribution obtained from the JETSET simulation of
quark-antiquark production and hadronization and does not exceed 10\%.  

The region $40<\chi_{4\pi 2\pi^0}^2<80$ is chosen as a control region for the
estimation of background from other ISR and non-ISR multihadron reactions. 

The signal region of Fig.~\ref{4pi2pi0_chi2_all} contains 35,499 data             
and 6,833 MC events, while for the control region the corresponding           
numbers are 8,421 and 672 respectively.                                  
\subsection{Background estimation}
\label{sec:background2}
The background subtraction procedure for the $2(\pipi\pi^0)$ final state is
identical  to that already described in Sec.~\ref{sec:background} for the $3(\pipi)$
final state,  using the $\chi_{4\pi 2\pi^0}^2$ distributions shown in
Fig.~\ref{4pi2pi0_chi2_all}.  The unshaded
histogram  of Fig.~\ref{4pi2pi0_babar} shows the $2(\pipi\pi^0)$
invariant  mass distribution
for  the signal region of Fig.~\ref{4pi2pi0_chi2_all}. 
The points with error bars show the ISR background
contribution obtained in the manner described from the control region of
Fig.~\ref{4pi2pi0_chi2_all}.  
The cross-hatched histogram in Fig.~\ref{4pi2pi0_babar} represents the non-ISR
background contribution obtained from the JETSET MC simulation.
Both backgrounds are relatively small at low mass (about 10\%), but the non-ISR
background accounts for about 20-25\% of the observed data at approximately
4\gevcc.  

Accounting for uncertainties in cross sections for background processes,
uncertainties in normalization of the control sample, and 
statistical fluctuations in the number of simulated events,
we estimate that this procedure for background subtraction results
in a systematic uncertainty of less than 3\% in the number of signal
events in the 1.6--3~\gevcc region of six-pion mass, but that the uncertainty            
increases to 5--10\% in the region above 3~\gevcc.

By selecting a ``background-free'' $2(\pipi\pi^0)\gamma$ sample with only four-charged
tracks and only five photons (about 5\% of events) we find that \chisq
distributions for data and MC have similar shapes. The ratio of MC
events selected by $\chi_{4\pi 2\pi^0}^2<1000$  and $\chi_{4\pi 2\pi^0}^2<40$
cuts is 1.14 reflecting soft photon radiation processes.
An estimate for this ratio can also be made directly from the data by
measuring  the relative number of $J/\psi$ events over continuum for the \chisq
regions  noted, yielding the ratio $1.08\pm0.04$, in good agreement with
the estimate  from simulation. The 6\% difference between the two estimates is taken
as the estimate of systematic error
for the $\chi_{4\pi 2\pi^0}^2<40$ selection.
\begin{figure}[tbh]
\includegraphics[width=0.9\linewidth]{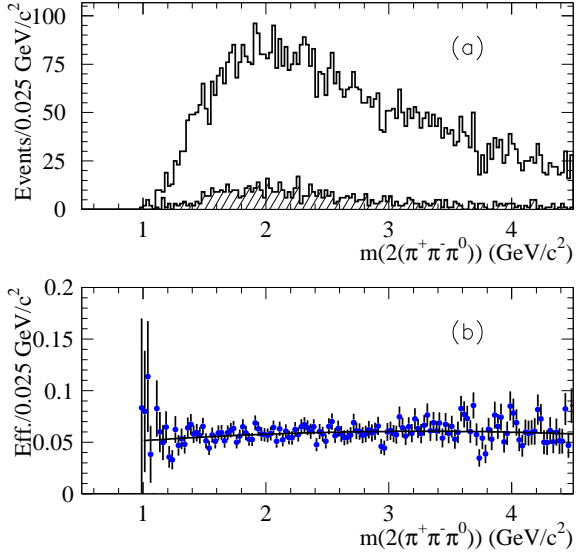}
\vspace{-0.4cm}
\caption{
(a) The six-pion (four charged and two neutral) mass distribution from
  MC simulation  for the
  signal and control (shaded) regions of Fig.~\ref{4pi2pi0_chi2_all}. (b) The mass
  dependence of the net reconstruction and selection efficiency
  obtained from simulation.
}
\label{mc_acc2}
\end{figure} 
\begin{figure}[tbh]
\includegraphics[width=0.9\linewidth]{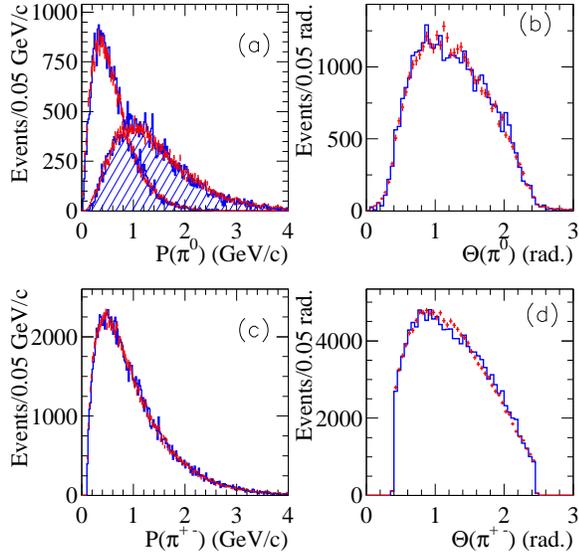}
\vspace{-0.4cm}
\caption{
Comparison of different kinematic parameters for $2(\pipi\pi^0)$ events
for data (points) and MC
simulation (histograms): (a) momentum distributions for the fast (shaded) and slow
(unshaded) $\pi^0$s from MC compared with the data; (b) combined polar 
angle distribution for the $\pi^0$s; (c) combined momentum
distribution for the two charged pions; (d) combined
polar angle distribution for the two charged pions.
}
\label{mom_theta}
\end{figure} 
\subsection{Pion-finding efficiency}
\label{sec:tracking2}
The charged-pion tracking inefficiency is corrected by applying a
$+(3\pm2)\%$ correction to
the number of observed $2(\pipi\pi^0)$ ISR events, following the
prescription discussed  above in Sec.~\ref{sec:tracking}.

The difference in the $\pi^0$-finding efficiencies between data and MC
 events has  been studied
 previously using the  $\pipi\pi^0\gamma$  reaction
~\cite{isr3pi}. To extend this  study to
 the case where  there are two $\pi^0$s in the final state, a high
 statistics sample  of  ISR-produced $\omega\pi^0\to\pipi\ppz$  events is
 selected using a 1C  fit that ignores the $\pi^0$  from the  $\omega$ decay. The
 $\pi^0$-finding  efficiency is then computed by comparing the number of
 events in the $\omega$  peak where the $\pi^0$ is found to the number of events
 where it is not.  By comparing data and MC results, it is found that
 the $\pi^0$ efficiency for simulation is 2.8$\pm$1.0$\pm$1.0\% higher where 
the systematic error comes mostly from the background subtraction procedure.
Assuming no correlation in $\pi^0$-finding efficiency for two $\pi^0$s
 in the event we apply
+5.6\% overall correction to which we assign a systematic error of 3\%.

\subsection{Detection efficiency from simulation}
The detection efficiency is determined in the same manner described in 
section~\ref{sec:eff}.
The simulated $2(\pipi\pi^0)$ invariant-mass
distributions after selection are shown in 
Fig.~\ref{mc_acc2}(a) for the signal and
control (shaded histogram) regions. 
The mass dependence of the detection
efficiency is obtained by dividing the number of reconstructed MC
events in each 25~\mevcc mass interval by the number generated in          
this same interval. The result is shown in
Fig.~\ref{mc_acc2}(b)
and demonstrates practically uniform efficiency versus mass.
This efficiency 
estimate takes into account the geometrical acceptance of the detector 
for the final-state photon and the charged and neutral pions, the
inefficiency of 
several detector subsystems, and event loss due to additional
soft-photon  emission from the initial and final states.

As mentioned in Sec.~\ref{sec:babar}, the model used in the MC simulation 
for $2(\pipi\pi^0)$ is pure phase space.
In general, this model describes the
distributions of all the kinematic variables characterizing the
final state well, as demonstrated in Fig.~\ref{mom_theta}.
The uniform angular acceptance in the rest 
frame of the final state pions makes the detection efficiency quite insensitive to
the presence of intermediate resonance structures. 
This feature has also been demonstrated in the six charged pion
analysis discussed  above and in our earlier study of four charged
pions~\cite{isr4pi} , where  MC simulations either with intermediate resonances
or with phase  space differ by no more than 3\% in detection
efficiency. We include  this 3\% here as a conservative estimate of
the  systematic  error for the model
dependence of the $2(\pipi\pi^0)$ detection efficiency.
\begin{figure}[tbh]
\includegraphics[width=0.9\linewidth]{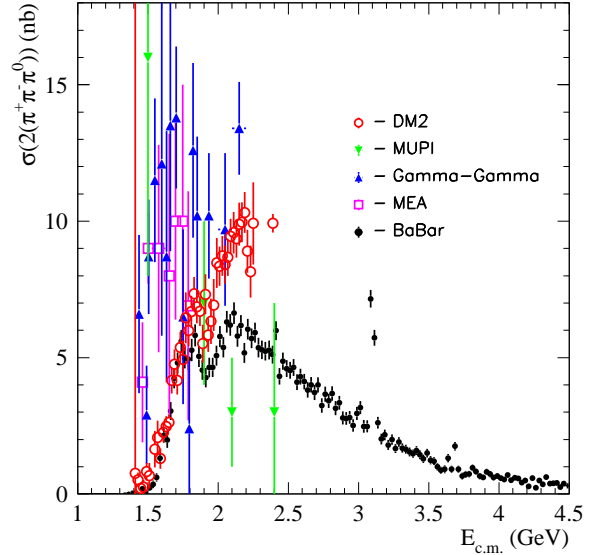}
\vspace{-0.5cm}
\caption{
The \epem c.m.\@ energy dependence of the $2(\pipi\pi^0)$ cross section
 measured with ISR data at \babar\ compared with the 
direct \epem measurements by detectors at ADONE and DCI. 
Only statistical errors are shown.
}
\label{4pi2pi0_ee_babar}
\end{figure} 
\subsection{\boldmath Cross section for $\epem\to 2(\pipi\pi^0)$}
\label{sec:4pi2pi0xs}
The \epem energy dependent cross section for the $2(\pipi\pi^0)$ final state can then
be calculated from
\begin{equation}
  \sigma(4\pi 2\pi^0)(E_{\rm c.m.})
  = \frac{dN_{4\pi 2\pi^0\gamma}(E_{\rm c.m.})}
         {d{\cal L}(E_{\rm c.m.})\cdot\epsilon_{4\pi 2\pi^0}^{\rm corr}
          \cdot\epsilon_{4\pi 2\pi^0}^{\rm MC}(E_{\rm c.m.})}\ ,
\end{equation}
where $m_{\rm inv}^{4\pi 2\pi^0} \equiv E_{\rm c.m.}$ with $m_{\rm
  inv}^{4\pi 2\pi^0}$ the invariant mass of 
the six-pion system; $dN_{4\pi 2\pi^0\gamma}$ is the number of selected
six-pion events after background subtraction in the interval $dE_{\rm c.m.}$ and
$\epsilon_{4\pi 2\pi^0}^{\rm MC}(E_{\rm c.m.})$ is the corresponding detection
efficiency obtained from the MC simulation. The factor
$\epsilon_{4\pi 2\pi^0}^{\rm corr}$ takes into account the 
differences between data and MC in tracking
and $\pi^0$ efficiencies, as
discussed in Sec.~\ref{sec:tracking2}.

\begin{table*}
\caption{Summary of the $\ep\en\to 2(\pipi\pi^0)$ 
cross section measurement. Errors are statistical only.}
\label{4pi2pi0_tab}
\begin{ruledtabular}
\hspace{-2. cm}
\begin{tabular}{ c c c c c c c c }
$E_{\rm c.m.}$ (GeV) & $\sigma$ (nb)  
& $E_{\rm c.m.}$ (GeV) & $\sigma$ (nb) 
& $E_{\rm c.m.}$ (GeV) & $\sigma$ (nb) 
& $E_{\rm c.m.}$ (GeV) & $\sigma$ (nb)  
\\
\hline
 1.3125 &  0.00 $\pm$  0.04 & 2.1125 &  6.44 $\pm$  0.43 & 2.9125 &  2.46 $\pm$  0.26 & 3.7125 &  0.74 $\pm$  0.15 \\
 1.3375 &  0.00 $\pm$  0.05 & 2.1375 &  5.45 $\pm$  0.42 & 2.9375 &  2.49 $\pm$  0.26 & 3.7375 &  0.51 $\pm$  0.13 \\
 1.3625 &  0.04 $\pm$  0.06 & 2.1625 &  5.92 $\pm$  0.41 & 2.9625 &  2.16 $\pm$  0.26 & 3.7625 &  0.60 $\pm$  0.14 \\
 1.3875 &  0.06 $\pm$  0.07 & 2.1875 &  4.81 $\pm$  0.40 & 2.9875 &  2.64 $\pm$  0.26 & 3.7875 &  0.64 $\pm$  0.13 \\
 1.4125 &  0.09 $\pm$  0.08 & 2.2125 &  5.81 $\pm$  0.40 & 3.0125 &  2.88 $\pm$  0.26 & 3.8125 &  0.86 $\pm$  0.14 \\
 1.4375 &  0.00 $\pm$  0.06 & 2.2375 &  5.45 $\pm$  0.39 & 3.0375 &  2.01 $\pm$  0.26 & 3.8375 &  0.73 $\pm$  0.13 \\
 1.4625 &  0.19 $\pm$  0.09 & 2.2625 &  5.68 $\pm$  0.39 & 3.0625 &  2.17 $\pm$  0.24 & 3.8625 &  0.53 $\pm$  0.13 \\
 1.4875 &  0.04 $\pm$  0.09 & 2.2875 &  5.04 $\pm$  0.39 & 3.0875 &  6.94 $\pm$  0.34 & 3.8875 &  0.53 $\pm$  0.12 \\
 1.5125 &  0.18 $\pm$  0.13 & 2.3125 &  5.10 $\pm$  0.36 & 3.1125 &  5.48 $\pm$  0.31 & 3.9125 &  0.70 $\pm$  0.13 \\
 1.5375 &  0.30 $\pm$  0.15 & 2.3375 &  4.99 $\pm$  0.37 & 3.1375 &  2.41 $\pm$  0.23 & 3.9375 &  0.62 $\pm$  0.13 \\
 1.5625 &  0.58 $\pm$  0.16 & 2.3625 &  5.01 $\pm$  0.37 & 3.1625 &  1.77 $\pm$  0.22 & 3.9625 &  0.50 $\pm$  0.12 \\
 1.5875 &  1.29 $\pm$  0.22 & 2.3875 &  4.88 $\pm$  0.36 & 3.1875 &  2.01 $\pm$  0.21 & 3.9875 &  0.55 $\pm$  0.12 \\
 1.6125 &  2.10 $\pm$  0.28 & 2.4125 &  5.86 $\pm$  0.37 & 3.2125 &  1.63 $\pm$  0.20 & 4.0125 &  0.47 $\pm$  0.12 \\
 1.6375 &  1.72 $\pm$  0.32 & 2.4375 &  4.07 $\pm$  0.33 & 3.2375 &  1.79 $\pm$  0.21 & 4.0375 &  0.42 $\pm$  0.12 \\
 1.6625 &  2.88 $\pm$  0.35 & 2.4625 &  4.70 $\pm$  0.34 & 3.2625 &  1.52 $\pm$  0.18 & 4.0625 &  0.60 $\pm$  0.12 \\
 1.6875 &  4.12 $\pm$  0.39 & 2.4875 &  4.37 $\pm$  0.33 & 3.2875 &  1.77 $\pm$  0.19 & 4.0875 &  0.40 $\pm$  0.11 \\
 1.7125 &  4.67 $\pm$  0.41 & 2.5125 &  4.25 $\pm$  0.34 & 3.3125 &  1.54 $\pm$  0.19 & 4.1125 &  0.50 $\pm$  0.11 \\
 1.7375 &  5.42 $\pm$  0.45 & 2.5375 &  4.39 $\pm$  0.33 & 3.3375 &  1.50 $\pm$  0.18 & 4.1375 &  0.46 $\pm$  0.11 \\
 1.7625 &  4.86 $\pm$  0.41 & 2.5625 &  3.85 $\pm$  0.31 & 3.3625 &  1.40 $\pm$  0.18 & 4.1625 &  0.32 $\pm$  0.11 \\
 1.7875 &  4.65 $\pm$  0.43 & 2.5875 &  4.13 $\pm$  0.31 & 3.3875 &  1.38 $\pm$  0.17 & 4.1875 &  0.45 $\pm$  0.10 \\
 1.8125 &  4.97 $\pm$  0.44 & 2.6125 &  3.92 $\pm$  0.31 & 3.4125 &  1.46 $\pm$  0.17 & 4.2125 &  0.17 $\pm$  0.10 \\
 1.8375 &  5.65 $\pm$  0.44 & 2.6375 &  3.49 $\pm$  0.31 & 3.4375 &  1.27 $\pm$  0.18 & 4.2375 &  0.50 $\pm$  0.11 \\
 1.8625 &  4.86 $\pm$  0.41 & 2.6625 &  3.74 $\pm$  0.31 & 3.4625 &  1.14 $\pm$  0.17 & 4.2625 &  0.14 $\pm$  0.10 \\
 1.8875 &  4.37 $\pm$  0.39 & 2.6875 &  3.41 $\pm$  0.30 & 3.4875 &  1.35 $\pm$  0.17 & 4.2875 &  0.39 $\pm$  0.11 \\
 1.9125 &  4.05 $\pm$  0.38 & 2.7125 &  3.72 $\pm$  0.30 & 3.5125 &  1.06 $\pm$  0.17 & 4.3125 &  0.26 $\pm$  0.10 \\
 1.9375 &  4.44 $\pm$  0.39 & 2.7375 &  2.94 $\pm$  0.29 & 3.5375 &  1.07 $\pm$  0.16 & 4.3375 &  0.56 $\pm$  0.10 \\
 1.9625 &  4.48 $\pm$  0.38 & 2.7625 &  3.34 $\pm$  0.30 & 3.5625 &  0.87 $\pm$  0.15 & 4.3625 &  0.26 $\pm$  0.11 \\
 1.9875 &  4.85 $\pm$  0.40 & 2.7875 &  3.11 $\pm$  0.29 & 3.5875 &  0.70 $\pm$  0.15 & 4.3875 &  0.30 $\pm$  0.10 \\
 2.0125 &  5.59 $\pm$  0.41 & 2.8125 &  3.43 $\pm$  0.29 & 3.6125 &  0.70 $\pm$  0.16 & 4.4125 &  0.34 $\pm$  0.10 \\
 2.0375 &  5.15 $\pm$  0.40 & 2.8375 &  2.81 $\pm$  0.28 & 3.6375 &  1.18 $\pm$  0.16 & 4.4375 &  0.16 $\pm$  0.10 \\
 2.0625 &  6.12 $\pm$  0.42 & 2.8625 &  3.12 $\pm$  0.27 & 3.6625 &  0.74 $\pm$  0.15 & 4.4625 &  0.32 $\pm$  0.10 \\
 2.0875 &  5.97 $\pm$  0.42 & 2.8875 &  2.45 $\pm$  0.26 & 3.6875 &  1.65 $\pm$  0.17 & 4.4875 &  0.17 $\pm$  0.10 \\
\end{tabular}
\end{ruledtabular}
\end{table*}

The energy dependence of the cross section for the reaction
$\epem\to 2(\pipi\pi^0)$ after all corrections is shown in
Fig.~\ref{4pi2pi0_ee_babar}. It again shows a structure around 1.9~\gev, 
reaching a peak value of about 6~nb near 2.0~\gev.
The cross section for each 25~\mev interval is presented in
Table~\ref{4pi2pi0_tab}.  

Since $d{\cal L}$ (see eq.~\ref{isrlum1}) has been corrected for vacuum polarization 
and final-state photon emission, the cross section includes effects 
due to vacuum polarization and final-state soft-photon emission.
For $g-2$ calculations, vacuum polarization
contributions should be excluded from our data. 

The observed line shape is not purely Gaussian mainly due to soft-photon
radiation.  Once again, no unfolding of the resolution is performed
for the results shown here.  The coefficients in equation 4 appropriate
to the 2(pi+pi-pi0) case are $e1,..e5$=0.007,0.091,0.744,0.114,0.011.

\begin{table*}[tbh]
\caption{
Summary of systematic errors for the $\epem\to 2(\pipi\pi^0)$ cross
section measurement.
}
\label{error2_tab}
\begin{ruledtabular}
\begin{tabular}{l c l} 
Source & Correction applied & Systematic error\\
\hline
Luminosity from $\mu\mu\gamma$ &  -  &  $3\%$ \\
MC-data difference in $\chi^{2}<40$ signal region & $0\%$ & $6\%$\\ 
Background subtraction & - &  $5\%$ for $m_{2(\pipi\pi^0)}<3.0~\gevcc$\\
                       &   &  $10\%$ for $m_{2(\pipi\pi^0)}>3.0~\gevcc$ \\
MC-data difference in tracking efficiency & $+3\%$ & $2\%$ \\

MC-data difference in $\pi^0$ losses & $+5.6\%$ & $3\%$ \\

Radiative corrections accuracy & - & $1\%$ \\
Acceptance from MC (model-dependent) & - & $3\%$  \\
\hline
Total(assuming addition in quadrature and no correlations)      &
$+8.8\%$    & $10\%$ for $m_{2(\pipi\pi^0)}<3.0~\gevcc$ \\
     &   & $13\%$ for $m_{2(\pipi\pi^0)}>3.0~\gevcc$\\
\end{tabular}
\end{ruledtabular}
\end{table*}
\subsection{\boldmath Summary of systematic studies}
\label{sec:Systematics2}
The measured cross sections for the $2(\pipi\pi^0)$ final state, shown in
Fig.~\ref{4pi2pi0_ee_babar} and tabulated  in Table~\ref{4pi2pi0_tab}, include
only statistical errors.
The systematic 
errors discussed in previous sections are summarized in
Table~\ref{error2_tab},  along 
with the corrections that were applied to the measurements.
The two systematic corrections applied to the measured cross sections 
sum to +8.8\% with 10-13\% systematic uncertainty.
\subsection{\boldmath Physics results}
\label{Physics2}
The cross section for the $2(\pipi\pi^0)$ final state measured by
\babar\ (Fig.~\ref{4pi2pi0_ee_babar}) can be compared with
existing \epem measurements performed by the $\mu\pi$~\cite{comb},
Gamma-Gamma~\cite{gaga} and MEA~\cite{mea} detectors at the ADONE collider
and by the DM2~\cite{6pidm2} detector at DCI. 
The \babar\ measurement is much more precise and disagrees with
DM2. The latter have probably large systematic errors
due to normalization, not discussed in the original reference~\cite{6pidm2}
that mostly focuses on the confirmation of the dip at 1.9 GeV
previously observed in the six-charged-pion mode.

\begin{figure}[tbh]
\includegraphics[width=1.0\linewidth]{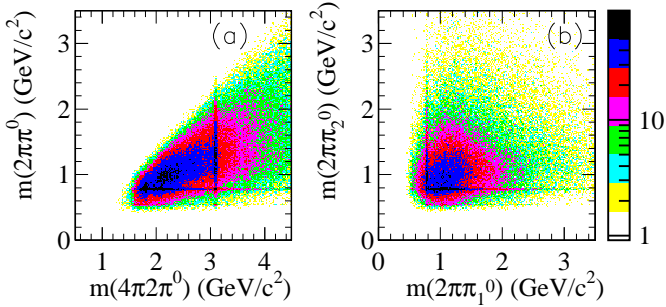}
\vspace{-0.4cm}
\caption{
Invariant mass scatter plots from data for; (a) the six pion
($2(\pipi\pi^0)$) final state versus  the neutral three-pion combinations
($\pipi\pi^0$); and (b)  one neutral three-pion combination with the lower
momentum $\pi^0$ versus the other with the higher momentum $\pi^0$.
}
\label{4pi2pi0_2-3pi}
\end{figure}
\begin{figure}[tbh]
\includegraphics[width=0.9\linewidth]{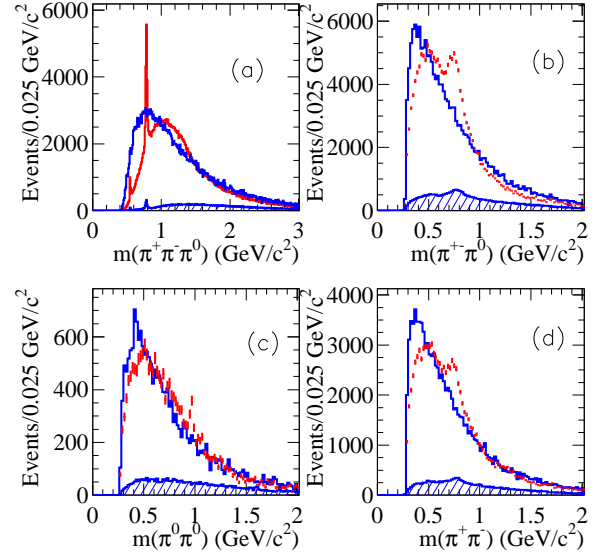}
\vspace{-0.4cm}
\caption{
Invariant mass distributions from data (points), simulation (unshaded
histograms), and  the non-ISR contributions obtained from JETSET
(shaded histograms)  for; (a) neutral three pion ($\pipi\pi^0$)
combinations; (b) charged  two pion ($\pi^\pm\pi^0$) combinations; (c) neutral
two body pi-zero ($\ppz$) combinations; and (d) neutral two body
charged pion  ($\pipi$) combinations.
}
\label{2pi_3pi_pro2}
\end{figure}
\begin{figure}[tbh]
\includegraphics[width=0.9\linewidth]{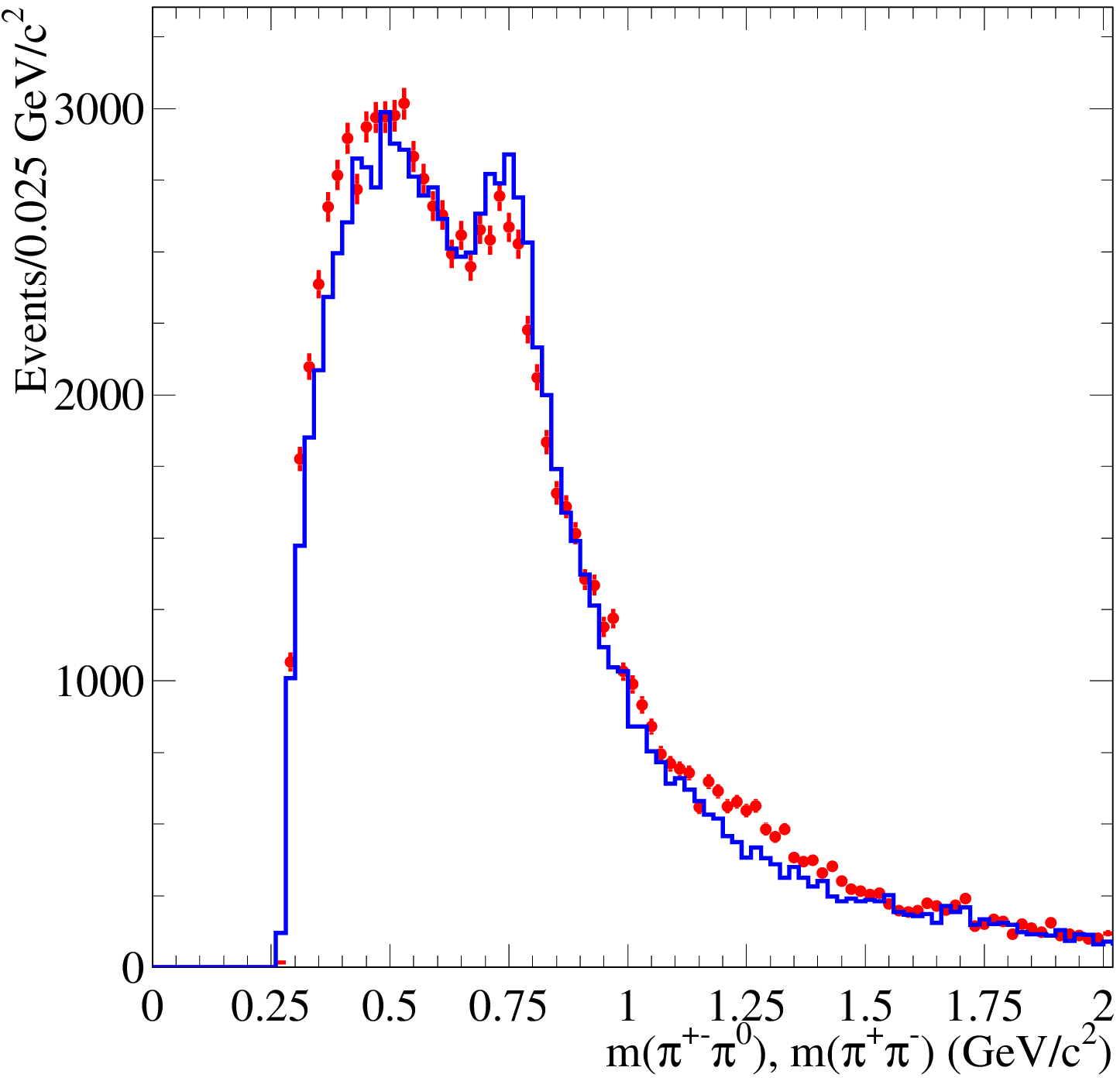}
\vspace{-0.4cm}
\caption{
The scaled $\pi^+\pi^-$ mass distribution (points, with errors,
multiplied by two)  compared with the summed $\pi^\pm\pi^0$ mass
distribution  (histogram).
}
\label{2pi_rho_pro2}
\end{figure}

Different mass combinations were studied in data and MC events to       
search for any structures or states not included in the simulation.

Figure~\ref{4pi2pi0_2-3pi} shows invariant mass scatter plots for; (a)
neutral three- versus six-pion combinations, and (b) neutral
$\pipi\pi^0$ combinations with the lower momentum $\pi^0$
versus three-pion with the higher momentum $\pi^0$.  The
$\omega(782)$ and  $\eta$ mesons are seen in
the three  body combinations. 
Figure~\ref{2pi_3pi_pro2}(a) shows the projection of the three-pion
invariant mass distribution  of Fig.~\ref{4pi2pi0_2-3pi} with clear $\omega(782)$
and $\eta$ signals. Figures ~\ref{2pi_3pi_pro2}(b,c,d)
show mass projections  for the two-pion combinations.
Backgrounds are subtracted from
the data points  shown using the $\chi^2$  control region, and the
non-ISR JETSET  simulation (shaded
histograms in Fig.~\ref{2pi_3pi_pro2}), as
described above.  The $\rho(770)$ meson is clearly seen in the
$\pi^{\pm}\pi^0$  and $\pi^+\pi^-$ combinations, while there is some
indication of a  small signal from $f_0 (980)$ in the $\ppz$ mass distribution.
The phase space MC simulation shown by unshaded histograms in
Fig.~\ref{2pi_3pi_pro2} does not include any of these structures.

Figure~\ref{2pi_rho_pro2} compares the $\pipi$ mass distribution
(multiplied by two) with  the summed $\pi^{\pm}\pi^0$ mass
distribution. The basic shapes  are quite similar, although there are
more than  twice as many charged as neutral  $\rho(770)$ mesons. The
neutral  $\pipi$  also has  a broad bump around 1.3~\gevcc relative to
the charged $\pi^{\pm}\pi^0$  distribution, perhaps indicating the
presence of  some intermediate $f_0 (1370)$ or $f_2 (1270)$ production.

No obvious structures are seen in the four or five pion combinations
(not shown). Though  the structures observed above are suggestive, a
partial wave analysis  would be needed to interpret the data more completely.

\begin{figure}[tbh]
\includegraphics[width=0.9\linewidth]{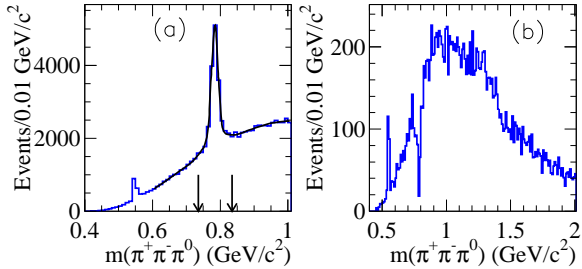}
\vspace{-0.4cm}
\caption{
The $\pipi\pi^0$ invariant mass distributions; (a) in the low mass
region; and (b) for the  second $\pipi\pi^0$ invariant mass
combination remaining in  those events where the first combination
lies within the  mass range indicated by the arrows in (a).
}
\label{omega_sel}
\end{figure}
\begin{figure}[tbh]
\includegraphics[width=0.9\linewidth]{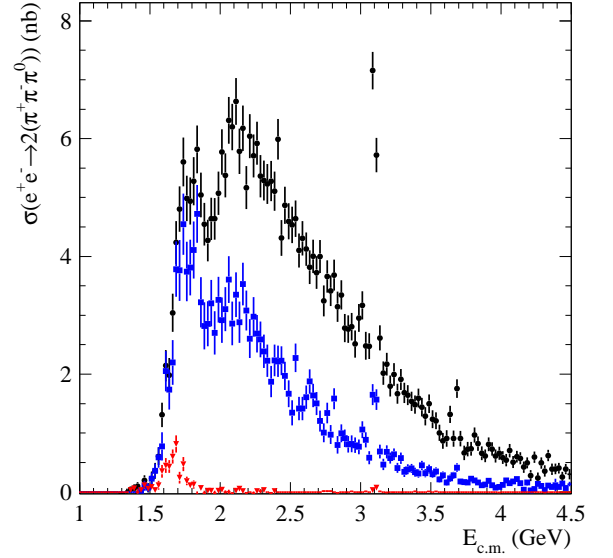}
\vspace{-0.4cm}
\caption{
The cross section versus E$_{c.m.}$ for all $2(\pipi\pi^0)$ events 
(circles), $\omega\pipi\pi^0$ events (squares), and $\omega\eta$
events (triangles).  
}
\label{omega_xs}
\end{figure}
\begin{figure}[tbh]
\includegraphics[width=0.9\linewidth]{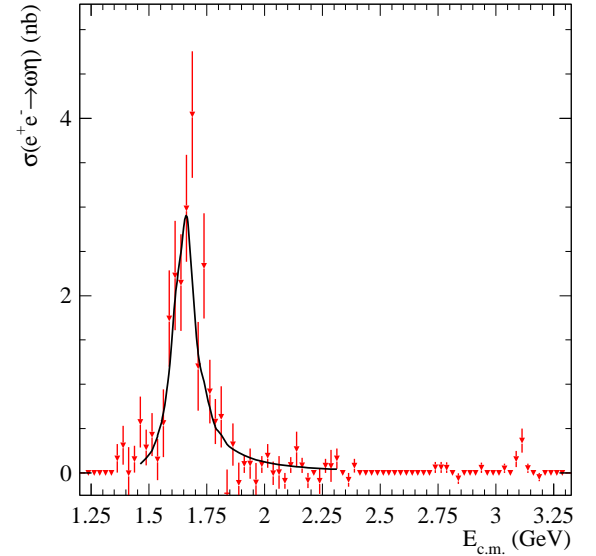}
\vspace{-0.4cm}
\caption{
The $\omega\eta$ cross section from the $2(\pipi\pi^0)$ event sample. The line is the
fit to the structure in the 1.6~\gev region described in the text.
}
\label{omega-eta_xs}
\end{figure}
\begin{figure}[tbh]
\includegraphics[width=0.9\linewidth]{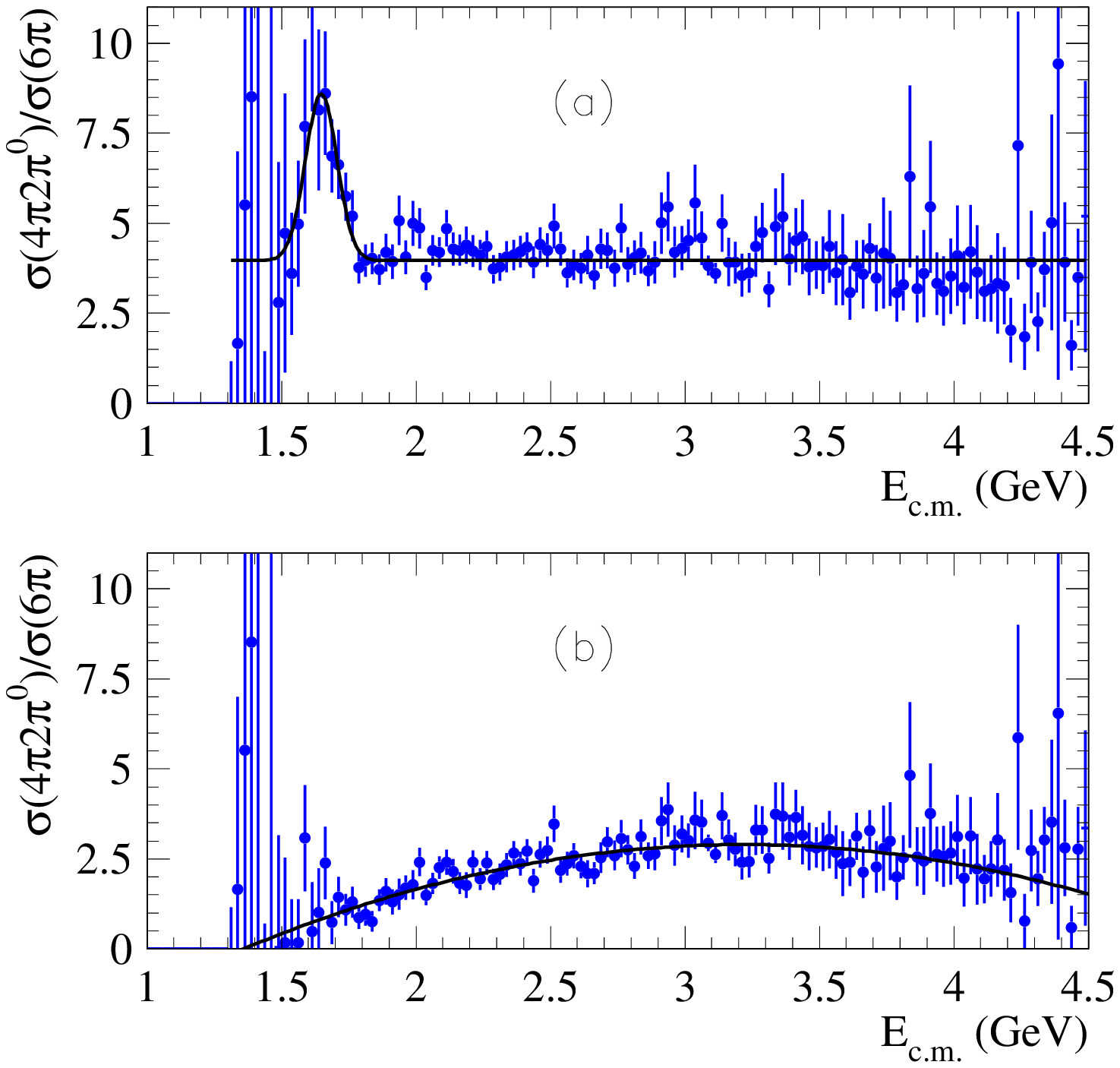}
\vspace{-0.4cm}
\caption{
Ratio of the $2(\pipi\pi^0)$  to $3(\pipi)$ cross sections for; (a)
selected events; (b) the  same events excluding the $\omega\pipi\pi^0$
contribution from  the $2(\pipi\pi^0)$ sample. The lines are fits described in the text.
}
\label{xs_ratio}
\end{figure}
Estimates for the $\omega(782)$  and $\eta$ contributions can be
obtained  using the $\pipi\pi^0$  mass distribution 
of Fig.~\ref{omega_sel}(a) which
shows an
expanded view of  Fig.~\ref{2pi_3pi_pro2}(a).  The procedure is to fit
each signal  with a two-Gaussian function representing the signal plus
a polynomial   background, yielding $9899\pm158$ and $770\pm40$
$\omega\pipi\pi^0$ and $\eta\pipi\pi^0$ events, respectively.

The energy dependence of the $\omega\pipi\pi^0$ cross section  is determined by
performing this  fit for each 25~\mevcc bin of the six-pion
($2(\pipi\pi^0)$) mass (Fig.~\ref{omega_xs}).  In addition to the broad
structure at  low mass, there is also a sharp structure in the $J/\psi$
region corresponding  to $170\pm24$ events decaying into the
$\omega\pipi\pi^0$ final state.  After correcting for efficiency and
normalizing  to the ISR luminosity, this yields the $\ep\en\to\omega\pipi\pi^0$
cross section shown by the squares in Fig.~\ref{omega_xs}. 

Six pion events that contain an $\omega$, defined by the arrows in
Fig.~\ref{omega_sel}(a),  also sometimes contain an $\eta$ as shown in
Fig.~\ref{omega_sel}(b).  After selecting these $\omega\eta$ events,
and subtracting  the background using the $\eta$ side bands, 
 we calculate the cross section for $\ep\en\to\omega\eta$ presented in
Fig.~\ref{omega-eta_xs}. The cross section is corrected for the decay rate
of $\omega$ and $\eta$ to $\pipi\pi^0$ taken from PDG~\cite{PDG}.
A prominent structure can be seen around 1.6 GeV, with a smaller peak
(from 13 events  over a background of less than 0.5) around 3.1 GeV
from $J/\psi\to\omega\eta$ decay.
The observed cross section is fitted with a resonance-type parameterization
\begin{eqnarray}
\sigma(\ep\en\to\omega\eta)=\frac{F(s)}{s^{3/2}}|A_{m}(s)|^2~~, \\
A_{m}(s)=\frac{m^{5/2}\Gamma_0\sqrt{\sigma_0/F(m^2)}}{s-m^2+i\sqrt{s}\Gamma_0}~~,
\end{eqnarray}
where $m$ is the mass, $\Gamma_0$ the width, and $\sigma_0$ the peak cross
section of this resonance production in $\ep\en$ collisions. $F(s)$ is
a phase space term equal to the cube of the $\omega$ (or $\eta$)
momentum in the $\omega\eta$ rest frame.
The fit gives $m = 1.645 \pm 0.008$~\gevcc, $\Gamma_0 = 0.114 \pm 0.014
$~\gev, and $\sigma_0 = 3.08 \pm 0.33$ nb. 

As discussed in section~\ref{sec:4pi2pi0xs},
the effect of mass resolution on the measured width of this structure
is small and not taken  into account. The mass value obtained is close
to the value $1670\pm30$~\gevcc listed in the PDG~\cite{PDG} for the
$\omega(1650)$, for which $\omega\eta$ final state has been seen by earlier
experiments, but the width seen here
is substantially  narrower than the $0.315 \pm 0.035$~\gev listed.  A
structure decaying  to $\omega\eta$ in this region might also
correspond to the $\phi(1680)$, whose  mass and width are listed in
the PDG as  $1.680 \pm 0.020$~\gevcc, and $0.150 \pm 0.050$~\gev,
respectively. However, no branching fraction of $\phi(1680)$ to
the $\omega\eta$ final state was reported in the previous experiments.

Figure~\ref{xs_ratio}(a)  shows the cross section ratio for
$2(\pipi\pi^0)$ to $3(\pipi)$, as given  by Figs.~\ref{omega_xs} and
~\ref{6pi_ee_babar},  respectively.  A good fit to this ratio for all
energy intervals, shown  by the curve in Fig.~\ref{xs_ratio}(a), is
obtained with a constant plus  a Gaussian in the 1.6~\gev region. The
ratio equals $3.98 \pm 0.06 \pm 0.41$ everywhere, except for the
region around 1.6~\gev, where it  reaches about 8 at peak. The
structure may be at least partially  explained by the presence of the
$\omega\eta$ structure in the $2(\pipi\pi^0)$ final state. When the
$\omega\pipi\pi^0$ contribution  is subtracted from the
$2(\pipi\pi^0)$ final state, the ratio  is no longer flat, as shown
by Fig.~\ref{xs_ratio}(b), and  the structure at low mass
disappears. A 2nd-order polynomial fits  the data well as shown by the
curve in Fig.~\ref{xs_ratio}(b).

\begin{figure}[tbh]
\includegraphics[width=0.9\linewidth]{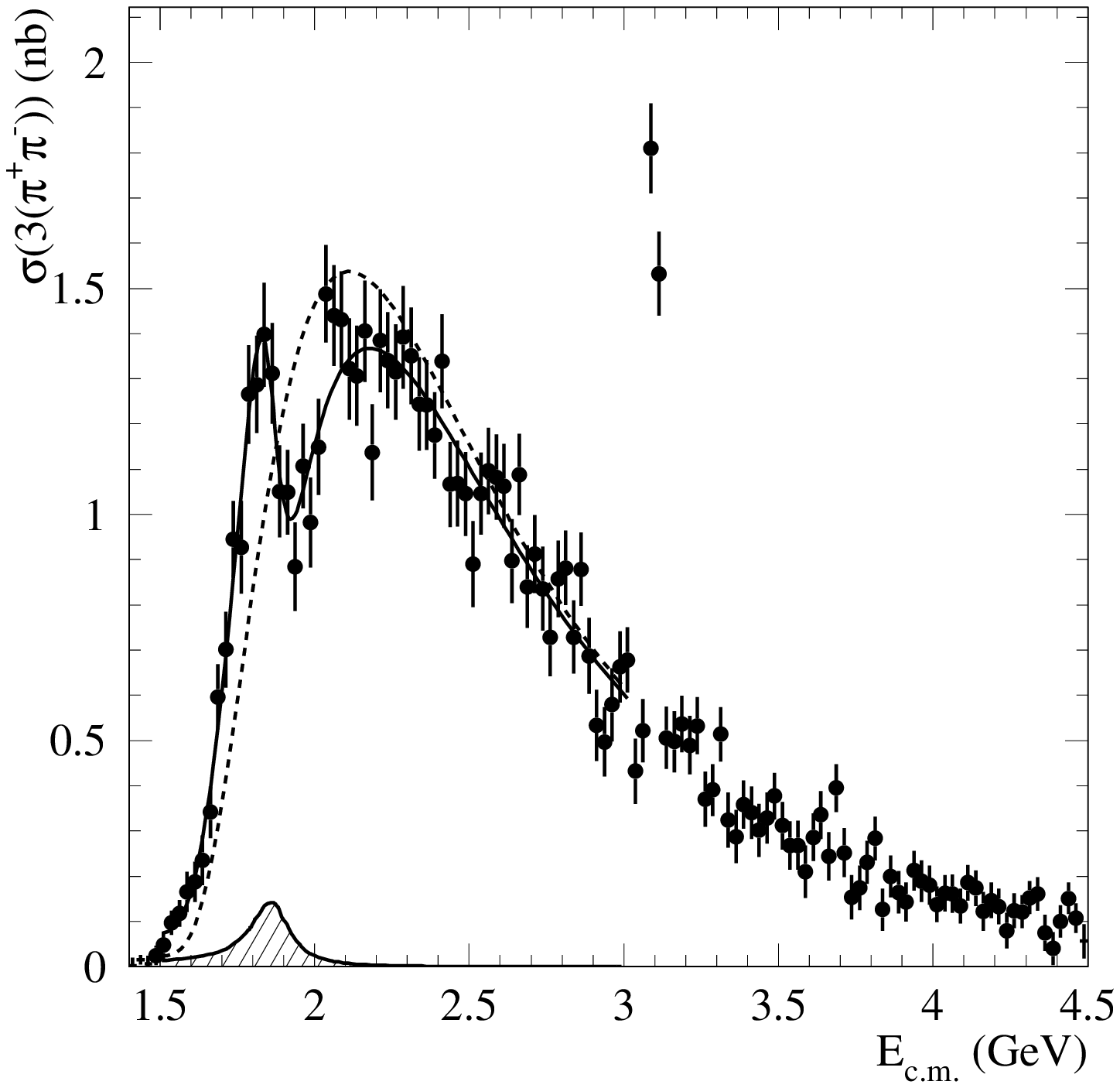}
\vspace{-0.4cm}
\caption{
The  $3(\pipi)$ cross section data, as given in
Fig.~\ref{6pi_ee_babar}, compared with the coherent  fit (solid line)
between resonance and continuum  terms described in the text. The
dashed line and the  shaded function show the individual incoherent
contributions  from continuum and resonance terms.
}
\label{dip_fit1}
\end{figure}
\begin{figure}[tbh]
\includegraphics[width=0.9\linewidth]{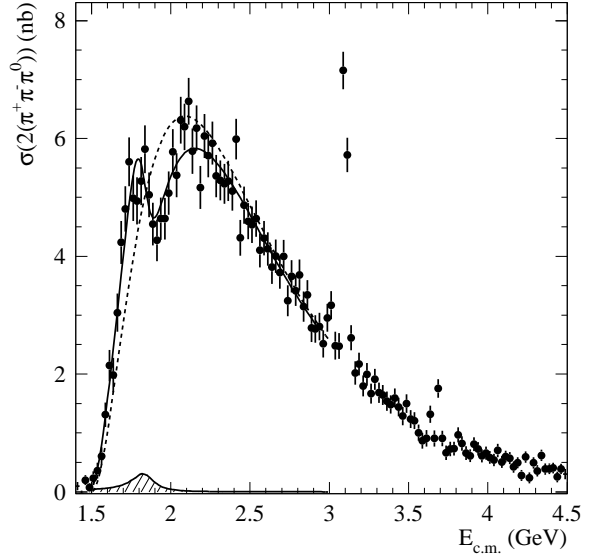}
\vspace{-0.4cm}
\caption{
The $2(\pipi\pi^0)$ cross section data, as given in
Fig.~\ref{omega_xs}, compared with the coherent  fit (solid line)
between resonance and  continuum terms described in the text. The
dashed line and the shaded  function show the individual incoherent
contributions  from continuum and resonance terms.
}
\label{dip_fit2}
\end{figure}
Figures~\ref{dip_fit1} and ~\ref{dip_fit2} compare the $3(\pipi)$ and
$2(\pipi\pi^0)$ cross section data,  respectively, with fits to the
model presented in Ref.~\cite{focus}.  The structures observed in
both channels around  1.9~\gev are not well described by a single
Breit-Wigner resonance, and may  result from rather complicated
physics, such as several vector  states decaying to the same mode. The
model~\cite{achasov} has  the form
\begin{equation}
\sigma_{6\pi} = \frac{4\pi\alpha^2}{s^{3/2}}\cdot
(\frac{g m^2 e^{i\phi}}{s-m^2+i\sqrt{s}\Gamma}+A_{cont})^2,
\end{equation}
where $m, \Gamma$ and $\phi$ are the mass, width and relative phase of the 
Breit-Wigner type amplitude, representing the structure, and 
$g$ is a  coupling
constant and $A_{cont}=c_0+c_1
\frac{e^{-b/(\sqrt{s}-m_0)}}{(\sqrt{s}-m_0)^{2-a}}$ is a Jacob-Slansky
amplitude~\cite{jacob} representing an amalgamation of broad
resonances with $c_0 , c_1 , a, b, m_0$ free parameters. 
The following ``resonance'' parameters are obtained for the structure:\\

$m_{6\pi}=1.88\pm0.03\gevcc; m_{4\pi 2\pi^0}=1.86\pm0.02\gevcc$, \\

$\Gamma_{6\pi} = 0.13 \pm 0.03~\gev; \Gamma_{4\pi 2\pi^0} = 0.16 \pm 0.02~\gev$,\\

$\phi_{6\pi} = 21 \pm 40^o~;~~~ \phi_{4\pi 2\pi^0} = -3 \pm 15^o~$.\\

The parameter values obtained seem to be essentially independent of
the final state  charge combination. These values may also be compared
with those obtained with a  similar model by the FOCUS
experiment~\cite{focus, focus1} $m = 1.91 \pm 0.01~\gevcc, \Gamma = 0.037 \pm
0.013~\gev, \phi = 10 \pm 30 ^o$ . The mass values  are consistent, but the widths
obtained by \babar\  are substantially larger.
%
\begin{figure}[tbh]
\includegraphics[width=0.9\linewidth]{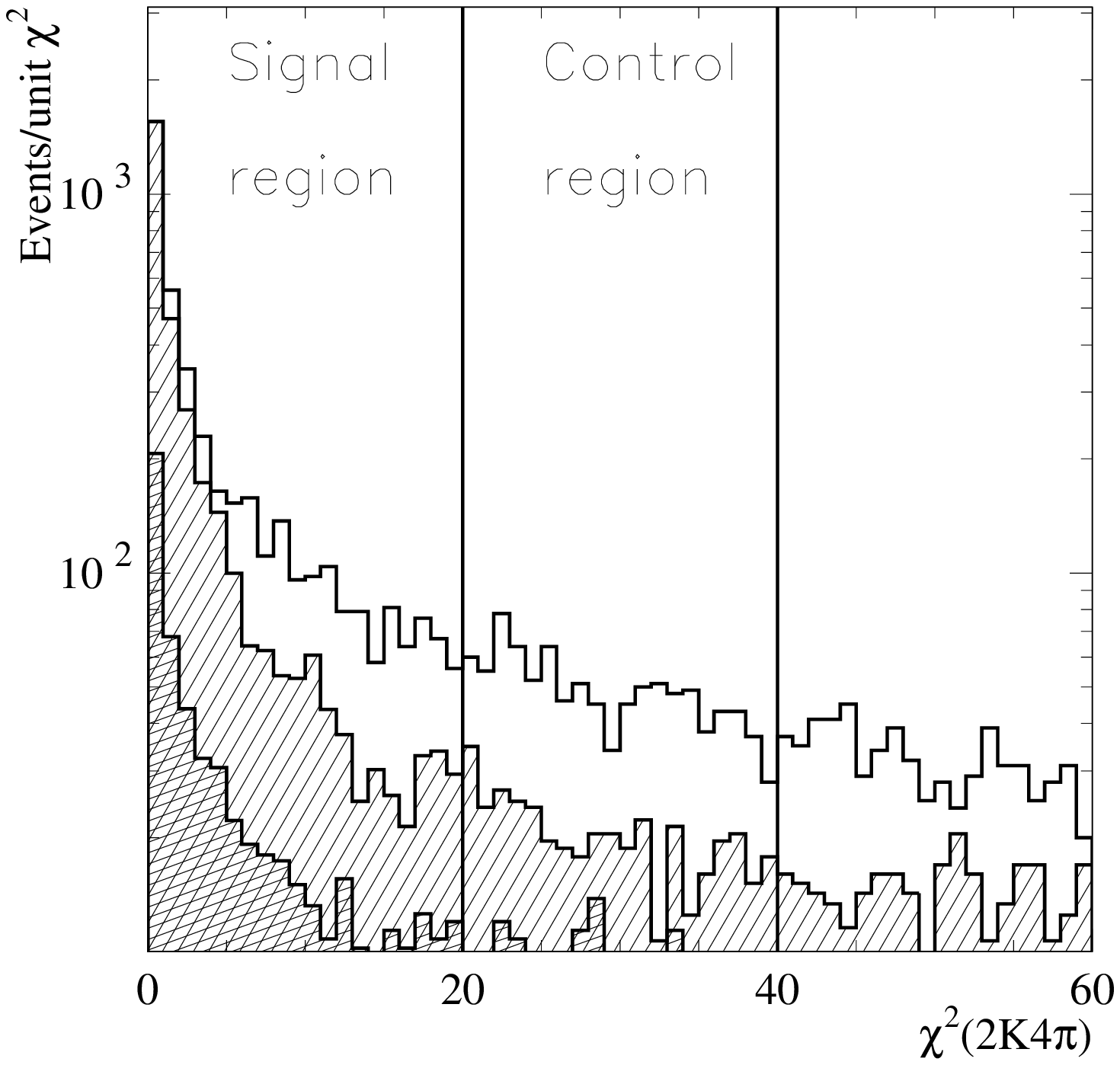}
\vspace{-0.4cm}
\caption{
The one-constraint \chisq distributions for data (unshaded histogram) and
       MC $K^+K^- 2(\pipi)\gamma$ simulation (shaded histogram) for six-charged-track events
       fitted to the $K^+ K^- 2(\pipi)$ hypothesis. The cross-hatched histogram
       is the estimated background contribution from non-ISR events
       obtained from JETSET. The signal and control regions are indicated.
}
\label{chi2_2k4pi}
\end{figure}
\begin{figure}[tbh]
\includegraphics[width=0.9\linewidth]{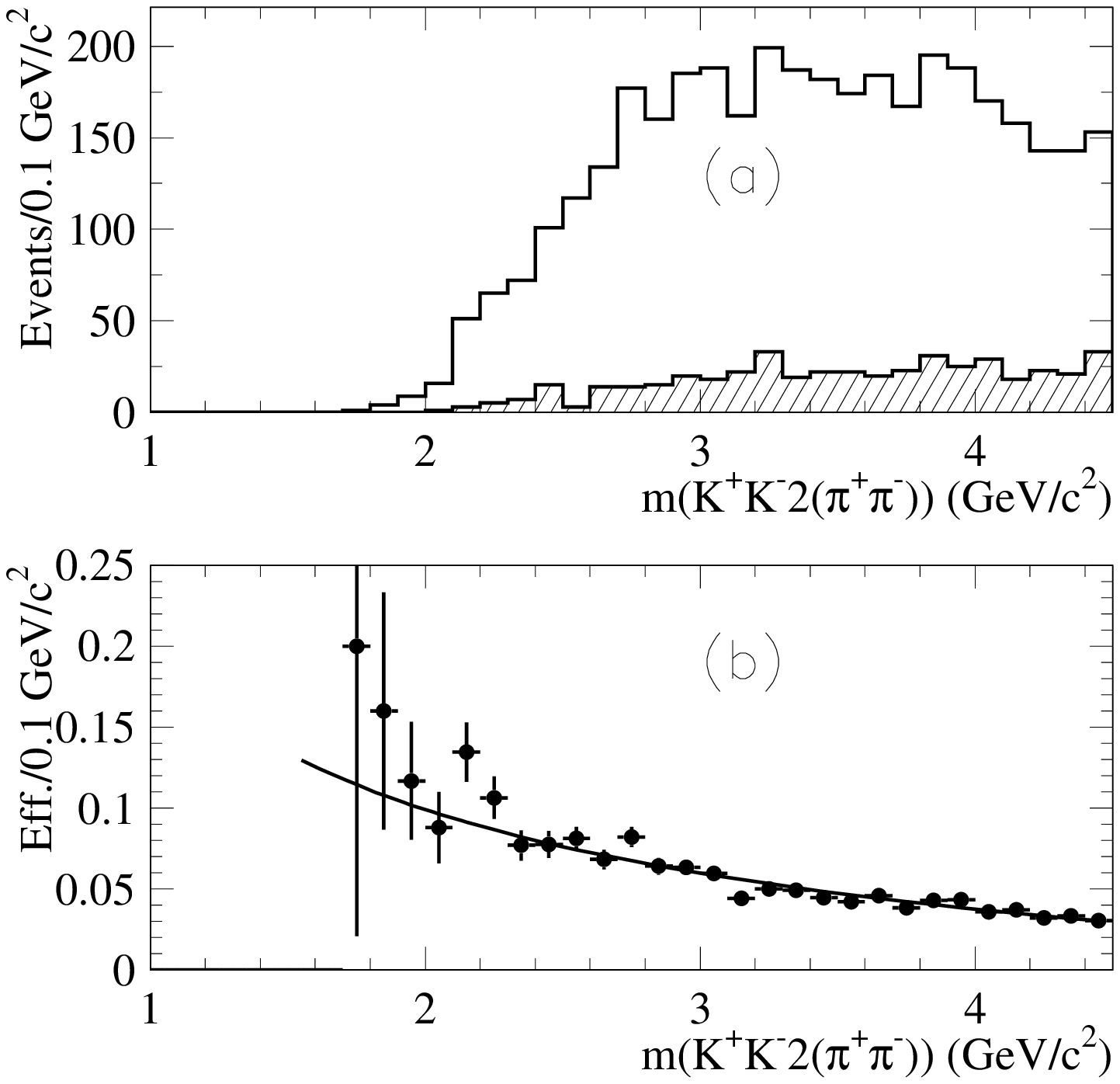}
\vspace{-0.4cm}
\caption{
(a) The $K^+ K^- 2(\pipi)$ mass distributions from MC simulation for the
  signal (unshaded) and control (shaded) regions of
  Fig.~\ref{chi2_2k4pi}. (b) The mass
  dependence of the net reconstruction and selection efficiency
  obtained from simulation. The curve is a fit described in the text.
}
\label{2k4pi_acc}
\end{figure} 
\section{\boldmath The $K^+K^- 2(\pipi)$ final state}\label{sec:2k4pi}
The constrained fit of the six-charged-track events to the hypothesis of
two oppositely charged kaons and four charged pions, where
at least one of the kaons has positive particle identification, allows
us to select this final state.
Figure~\ref{chi2_2k4pi} shows the \chisq
distributions for both data and simulation, where the simulation of the 
$K^+K^- 2(\pipi)$ reaction uses a phase space model
with a cross section energy dependence close to that which we observe 
experimentally, and ISR and FSR extra radiative processes are
included. Also shown is the estimated contribution from 
non-ISR events obtained by JETSET simulation.

Figure~\ref{2k4pi_acc}(a) presents the simulated mass distribution for
the $2K4\pi$ events. The mass dependence of the
efficiency, calculated as a ratio of selected  
to generated $2K4\pi$ MC events, is shown in Fig.~\ref{2k4pi_acc}(b).
The efficiency falls gradually from about 15\% at low mass, to about
3\% at 4.5~\gevcc. The data  are well represented by the 3rd-order
polynomial shown by  the curve in Fig.~\ref{2k4pi_acc}(b).
\begin{figure}[tbh]
\includegraphics[width=0.9\linewidth]{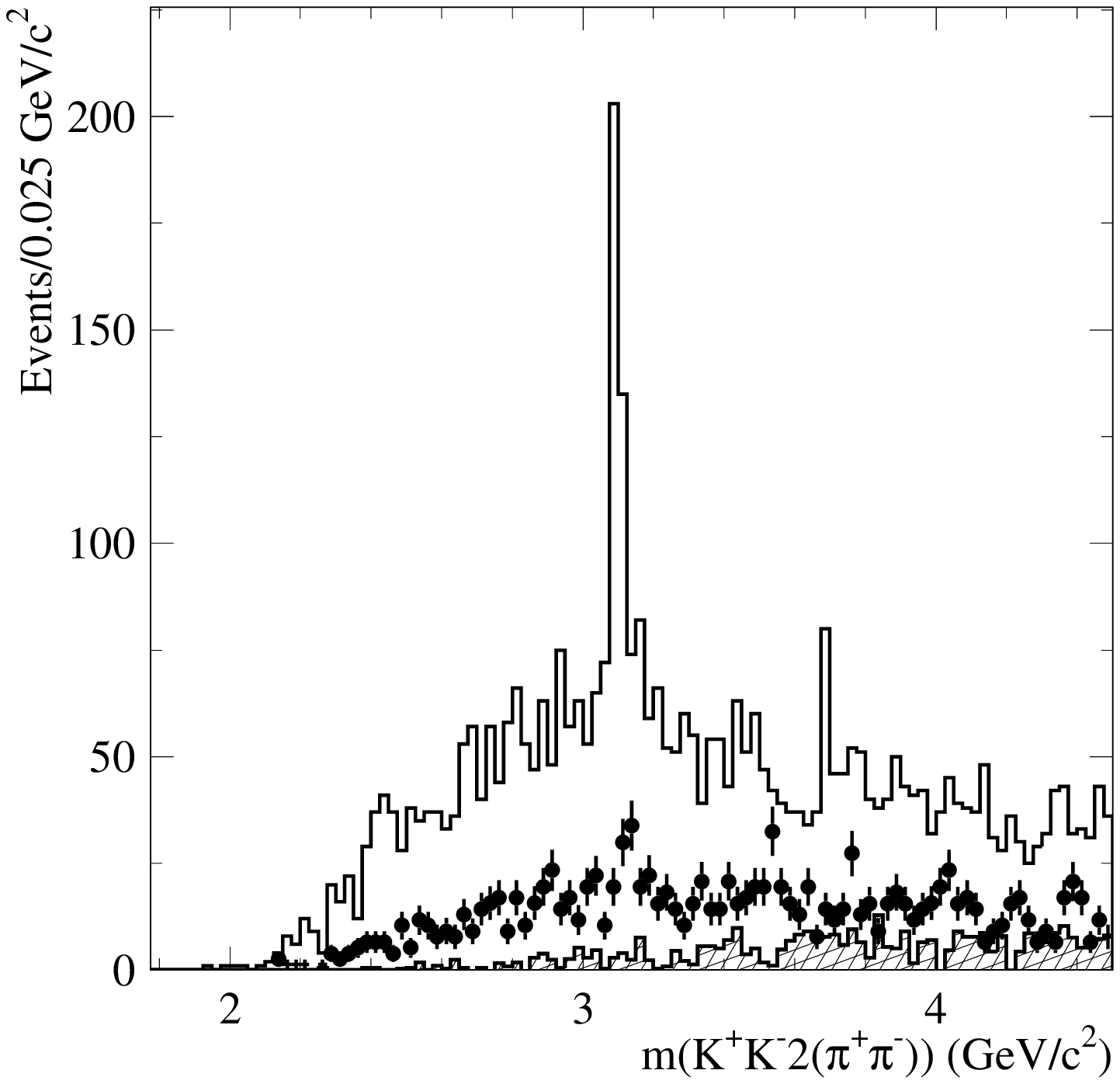}
\vspace{-0.4cm}
\caption{
The $K^+K^- 2(\pipi)$ invariant mass distribution (unshaded histogram)
for the  signal region of 
Fig.~\ref{chi2_2k4pi}, after subtracting mis-identified six charged
pion events as described in the text.  
 The points indicate the background estimated from
the difference between data and MC events for the control region of
Fig.~\ref{chi2_2k4pi}, normalized to the difference between data and MC
events in
the signal region of Fig.~\ref{chi2_2k4pi}. The cross-hatched histogram
corresponds to the non-ISR background of Fig.~\ref{chi2_2k4pi}.
}                          
\label{2k4pi}
\end{figure}
\begin{figure}[tbh]
\includegraphics[width=0.9\linewidth]{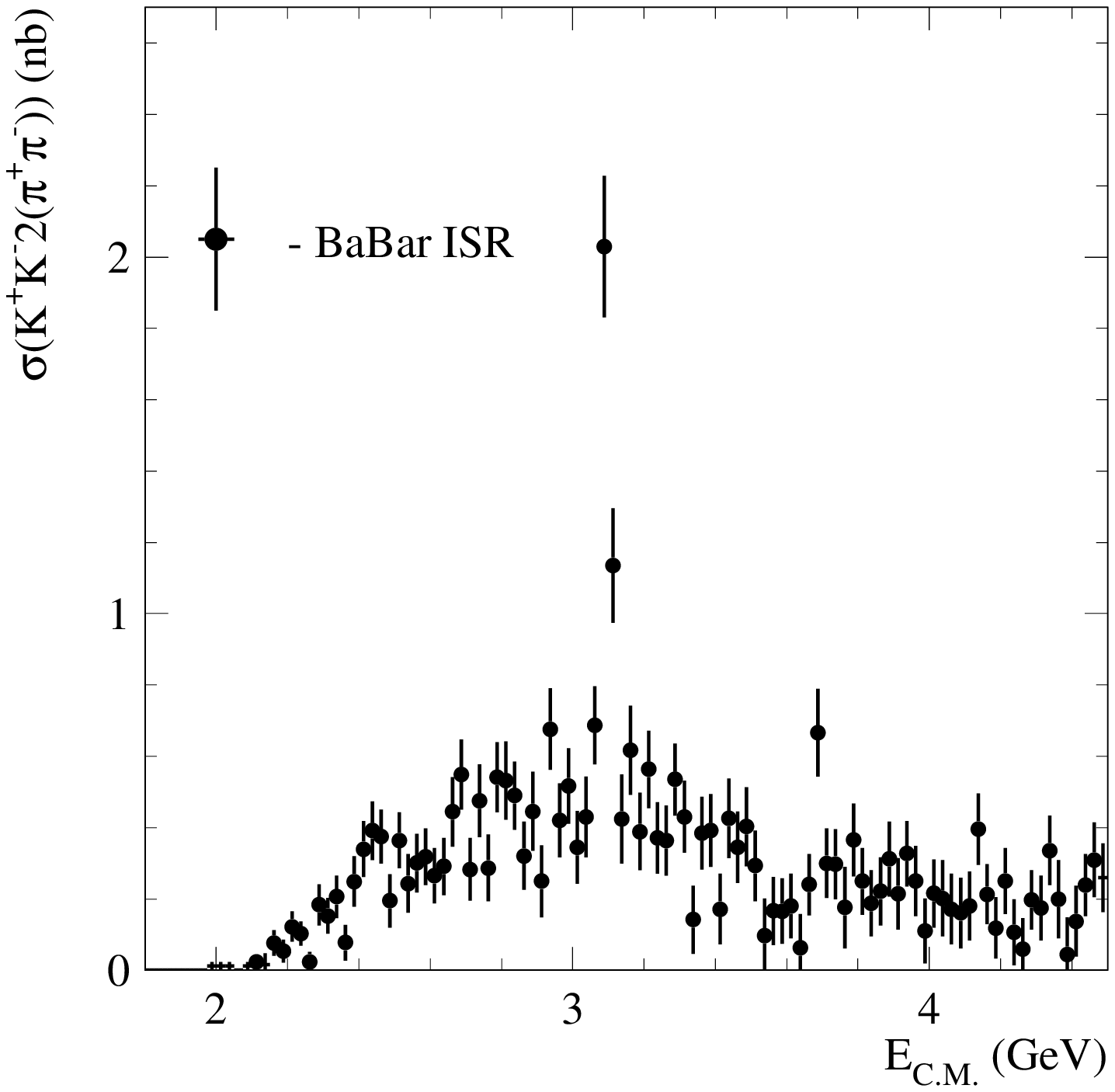}
\vspace{-0.4cm}
\caption{
The c.m.\@ energy dependence of the $\epem\to K^+K^- 2(\pipi)$
cross section obtained from ISR events at \babar. 
Only statistical errors are shown.
}
\label{2k4pi_xs}
\end{figure} 
After the selections $\chi^2_{2K4\pi}<20$, $\chi^2_{6\pi}>20$ are
imposed, we estimate from simulation  that about 0.6\% of the six
charged pion events remain in the  sample due to misidentification of
pions. Although the background  subtraction procedure, as described
above, using the control  region $20<\chi^2_{2K4\pi}<40$ is invoked,
background from pion  misidentification will not be subtracted by this
procedure. Thus, 0.6\% of the  six pion events shown in
Fig.~\ref{6pi_babar} are  subtracted, leading to a correction of about
3\% to  the total number of 
$K^+K^- 2(\pipi)$ events. 

Figure ~\ref{2k4pi} shows the $K^+K^- 2(\pipi)$  invariant mass
distributions for signal events,  selected as defined above,  as well
as for
events from the control region  in $\chi^2_{2K4\pi}$,  and for the
expected non-ISR  background from JETSET MC simulation. 
These latter two distributions are used
to subtract  background in the signal region. Clear $J/\psi$ and
$\psi(2S)$ signals are seen.

\begin{table*}
\caption{Summary of the  $\ep\en\to K^+ K^- 2(\pipi)$ 
cross section measurement. Errors are statistical only.}
\label{2k4pi_tab}
\begin{ruledtabular}
\begin{tabular}{ c c c c c c c c }
$E_{\rm c.m.}$ (GeV) & $\sigma$ (nb)  
& $E_{\rm c.m.}$ (GeV) & $\sigma$ (nb) 
& $E_{\rm c.m.}$ (GeV) & $\sigma$ (nb) 
& $E_{\rm c.m.}$ (GeV) & $\sigma$ (nb)  
\\
\hline

 2.0125 &  0.01 $\pm$  0.01 & 2.6375 &  0.29 $\pm$  0.08 & 3.2625 &  0.36 $\pm$  0.10 & 3.8875 &  0.31 $\pm$  0.11 \\
 2.0375 &  0.01 $\pm$  0.01 & 2.6625 &  0.44 $\pm$  0.10 & 3.2875 &  0.54 $\pm$  0.10 & 3.9125 &  0.21 $\pm$  0.10 \\
 2.0625 &  0.00 $\pm$  0.00 & 2.6875 &  0.55 $\pm$  0.10 & 3.3125 &  0.43 $\pm$  0.10 & 3.9375 &  0.33 $\pm$  0.09 \\
 2.0875 &  0.01 $\pm$  0.01 & 2.7125 &  0.28 $\pm$  0.09 & 3.3375 &  0.14 $\pm$  0.10 & 3.9625 &  0.25 $\pm$  0.10 \\
 2.1125 &  0.02 $\pm$  0.02 & 2.7375 &  0.47 $\pm$  0.10 & 3.3625 &  0.38 $\pm$  0.10 & 3.9875 &  0.11 $\pm$  0.09 \\
 2.1375 &  0.02 $\pm$  0.03 & 2.7625 &  0.29 $\pm$  0.09 & 3.3875 &  0.39 $\pm$  0.10 & 4.0125 &  0.22 $\pm$  0.09 \\
 2.1625 &  0.08 $\pm$  0.04 & 2.7875 &  0.54 $\pm$  0.10 & 3.4125 &  0.17 $\pm$  0.10 & 4.0375 &  0.20 $\pm$  0.11 \\
 2.1875 &  0.05 $\pm$  0.03 & 2.8125 &  0.53 $\pm$  0.11 & 3.4375 &  0.43 $\pm$  0.11 & 4.0625 &  0.17 $\pm$  0.10 \\
 2.2125 &  0.12 $\pm$  0.04 & 2.8375 &  0.49 $\pm$  0.10 & 3.4625 &  0.34 $\pm$  0.10 & 4.0875 &  0.16 $\pm$  0.10 \\
 2.2375 &  0.10 $\pm$  0.03 & 2.8625 &  0.32 $\pm$  0.10 & 3.4875 &  0.40 $\pm$  0.11 & 4.1125 &  0.18 $\pm$  0.10 \\
 2.2625 &  0.02 $\pm$  0.03 & 2.8875 &  0.45 $\pm$  0.11 & 3.5125 &  0.29 $\pm$  0.10 & 4.1375 &  0.40 $\pm$  0.10 \\
 2.2875 &  0.18 $\pm$  0.06 & 2.9125 &  0.25 $\pm$  0.10 & 3.5375 &  0.10 $\pm$  0.11 & 4.1625 &  0.21 $\pm$  0.08 \\
 2.3125 &  0.15 $\pm$  0.05 & 2.9375 &  0.68 $\pm$  0.11 & 3.5625 &  0.17 $\pm$  0.10 & 4.1875 &  0.12 $\pm$  0.09 \\
 2.3375 &  0.21 $\pm$  0.06 & 2.9625 &  0.42 $\pm$  0.10 & 3.5875 &  0.17 $\pm$  0.09 & 4.2125 &  0.25 $\pm$  0.09 \\
 2.3625 &  0.08 $\pm$  0.05 & 2.9875 &  0.52 $\pm$  0.11 & 3.6125 &  0.18 $\pm$  0.09 & 4.2375 &  0.11 $\pm$  0.09 \\
 2.3875 &  0.25 $\pm$  0.07 & 3.0125 &  0.34 $\pm$  0.10 & 3.6375 &  0.06 $\pm$  0.09 & 4.2625 &  0.06 $\pm$  0.09 \\
 2.4125 &  0.34 $\pm$  0.08 & 3.0375 &  0.43 $\pm$  0.11 & 3.6625 &  0.24 $\pm$  0.09 & 4.2875 &  0.20 $\pm$  0.08 \\
 2.4375 &  0.39 $\pm$  0.08 & 3.0625 &  0.69 $\pm$  0.11 & 3.6875 &  0.67 $\pm$  0.12 & 4.3125 &  0.17 $\pm$  0.09 \\
 2.4625 &  0.38 $\pm$  0.08 & 3.0875 &  2.03 $\pm$  0.20 & 3.7125 &  0.30 $\pm$  0.10 & 4.3375 &  0.34 $\pm$  0.10 \\
 2.4875 &  0.19 $\pm$  0.07 & 3.1125 &  1.14 $\pm$  0.16 & 3.7375 &  0.30 $\pm$  0.10 & 4.3625 &  0.20 $\pm$  0.11 \\
 2.5125 &  0.36 $\pm$  0.08 & 3.1375 &  0.42 $\pm$  0.13 & 3.7625 &  0.18 $\pm$  0.11 & 4.3875 &  0.04 $\pm$  0.10 \\
 2.5375 &  0.24 $\pm$  0.08 & 3.1625 &  0.62 $\pm$  0.12 & 3.7875 &  0.37 $\pm$  0.10 & 4.4125 &  0.14 $\pm$  0.10 \\
 2.5625 &  0.30 $\pm$  0.08 & 3.1875 &  0.39 $\pm$  0.11 & 3.8125 &  0.25 $\pm$  0.09 & 4.4375 &  0.24 $\pm$  0.09 \\
 2.5875 &  0.32 $\pm$  0.08 & 3.2125 &  0.56 $\pm$  0.11 & 3.8375 &  0.19 $\pm$  0.09 & 4.4625 &  0.31 $\pm$  0.11 \\
 2.6125 &  0.26 $\pm$  0.08 & 3.2375 &  0.37 $\pm$  0.10 & 3.8625 &  0.22 $\pm$  0.10 & 4.4875 &  0.26 $\pm$  0.10 \\
\end{tabular}
\end{ruledtabular}
\end{table*}

Using the number of observed events, efficiency, and ISR luminosity, we
obtain the $\epem\to
K^+K^- 2(\pipi)$ cross section shown in
Fig.~\ref{2k4pi_xs}. 
No measurements are available from earlier experiments.
Table~\ref{2k4pi_tab}
presents the cross section in 25~\mev bins.
The systematic errors are dominated by the uncertainty in the acceptance 
simulation (10\%), by the uncertainty in the background subtraction (5\%), and 
by the difference between the kaon identification 
efficiencies for data and MC events (up to 
2\% per track), and are estimated to be about 15\%. 

\begin{figure}
\includegraphics[width=0.95\linewidth,height=0.9\linewidth]{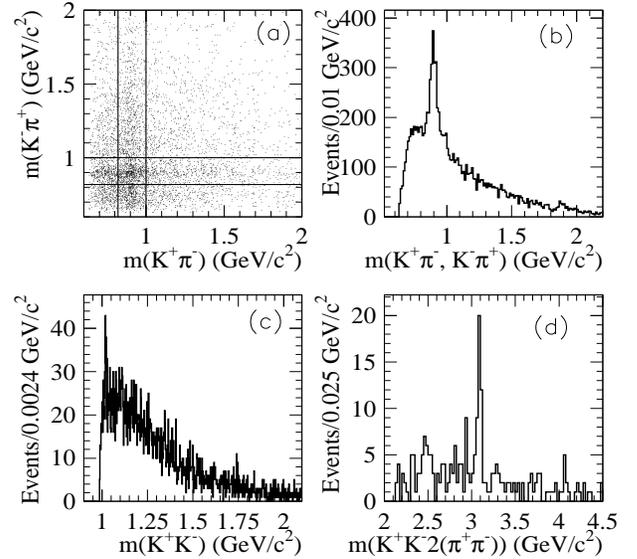}
\vspace{-0.4cm}
\caption{
Invariant mass plots for various selections of data from the
$K^+K^- 2(\pipi)$ sample for:  
(a) the scatter-plot of
the $K^+\pi^-$ and $K^-\pi^+$ invariant mass values; 
(b) the $K^+\pi^-$ or $K^-\pi^+$ mass projection of (a);
(c) the mass distribution for $K^+K^-$;
(d) the $K^+K^- 2(\pipi)$ mass distribution for events around
the $\phi(1020)$ peak from (c).
}
\label{2k4pi_mass1}
\end{figure}

Figures~\ref{2k4pi_mass1}(a,b) show the $K\pi$ mass combinations.
Production of $K\pi$ pairs is dominated by the $K^{*0}(892)$
clearly seen in Fig.~\ref{2k4pi_mass1}(a) and the projection plot in
Fig.~\ref{2k4pi_mass1}(b). Some small structure is also seen at about 1.9~\gevcc. 
Figure~\ref{2k4pi_mass1}(c) shows the $K^+K^-$ mass distribution. No
structures except a small signal from $\phi(1020)$ are seen. These
events correspond to  $J/\psi\to\phi (1020) 2(\pipi)$ decay. This is demonstrated 
in Fig.~\ref{2k4pi_mass1}(d) where events from the 1.01--1.03~\gevcc mass interval of 
Fig.~\ref{2k4pi_mass1}(c) are shown. We find $35\pm7$ events in the $J/\psi$ peak 
from the above decay chain.

\section{\boldmath The $J/\psi$ region}
\begin{figure}[tbh]
\includegraphics[width=0.9\linewidth]{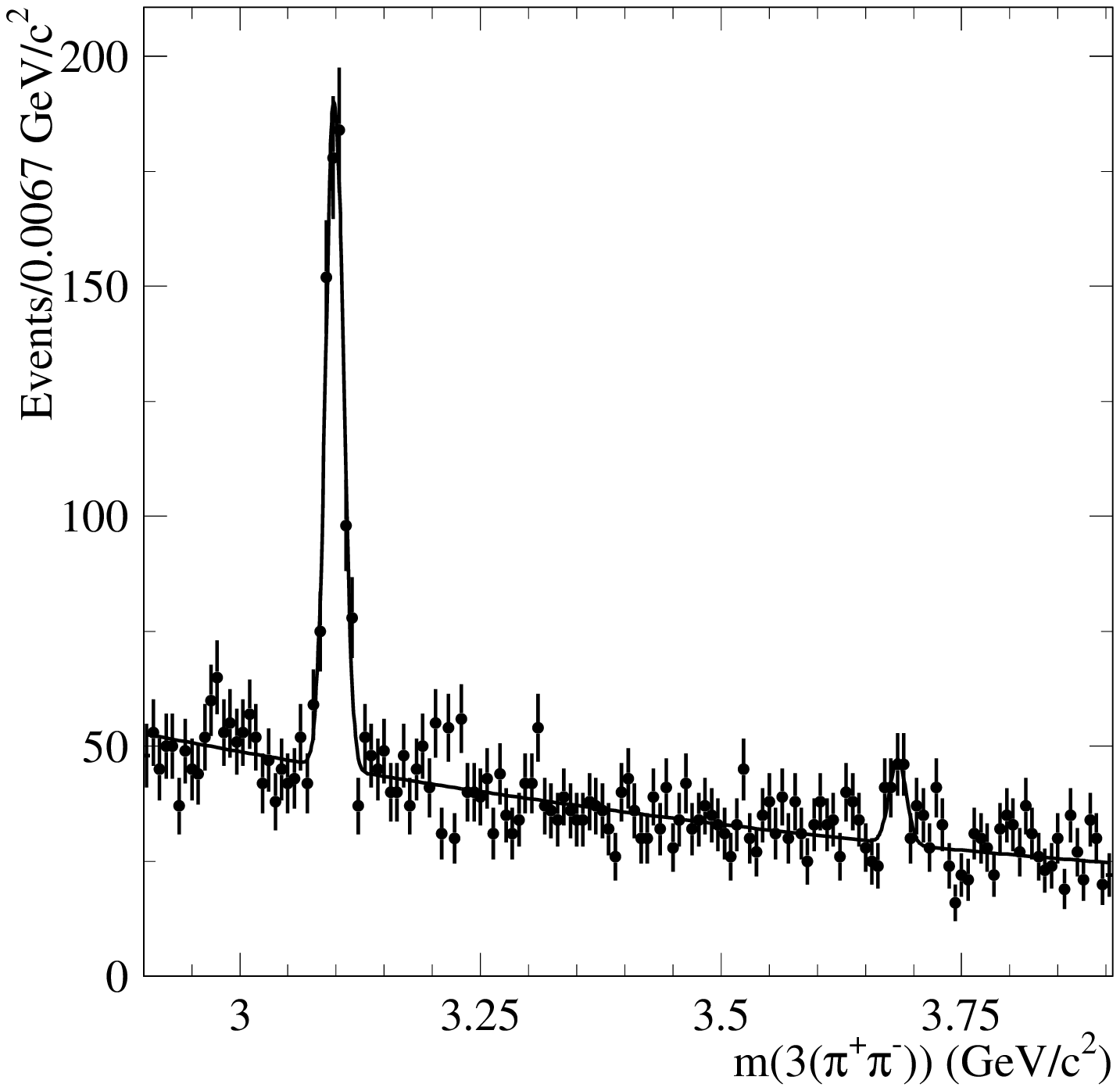}
\vspace{-0.4cm}
\caption{
The $3(\pipi)$ mass distribution for ISR-produced
$\epem\to 3(\pipi)$ events  
in the mass region around the $J/\psi$ and $\psi(2S)$;
there are clear signals at the $J/\psi$ and 
        $\psi(2S)$ mass positions. 
}
\label{jpsi}
\end{figure}

Figure~\ref{jpsi} shows an expanded view of the $J/\psi$ mass region 
in Fig.~\ref{6pi_babar} for the six-pion data sample with no background subtraction. 
The signals from $J/\psi\to 3(\pipi)$ and $\psi(2S)\to 3(\pipi)$ are clearly seen. 
\begin{figure}[tbh]
\includegraphics[width=0.9\linewidth]{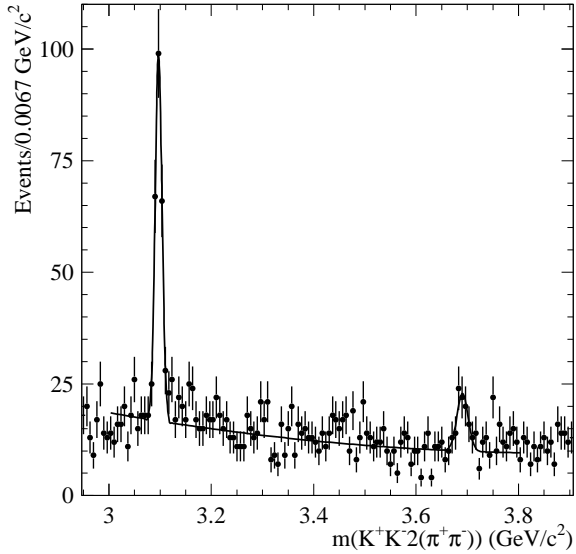}
\vspace{-0.4cm}
\caption{
The $K^+K^- 2(\pipi)$ mass 
distribution for ISR-produced events
in the mass region around the $J/\psi$ and $\psi(2S)$;
there are clear signals at the $J/\psi$ and $\psi(2S)$ mass positions.
The latter is dominated by $\psi(2S)\to J/\psi\pipi$, with 
        $J/\psi\to K^+K^-\pipi$.
}
\label{jpsi_2k4pi}
\end{figure}
\begin{figure}[tbh]
\includegraphics[width=0.9\linewidth]{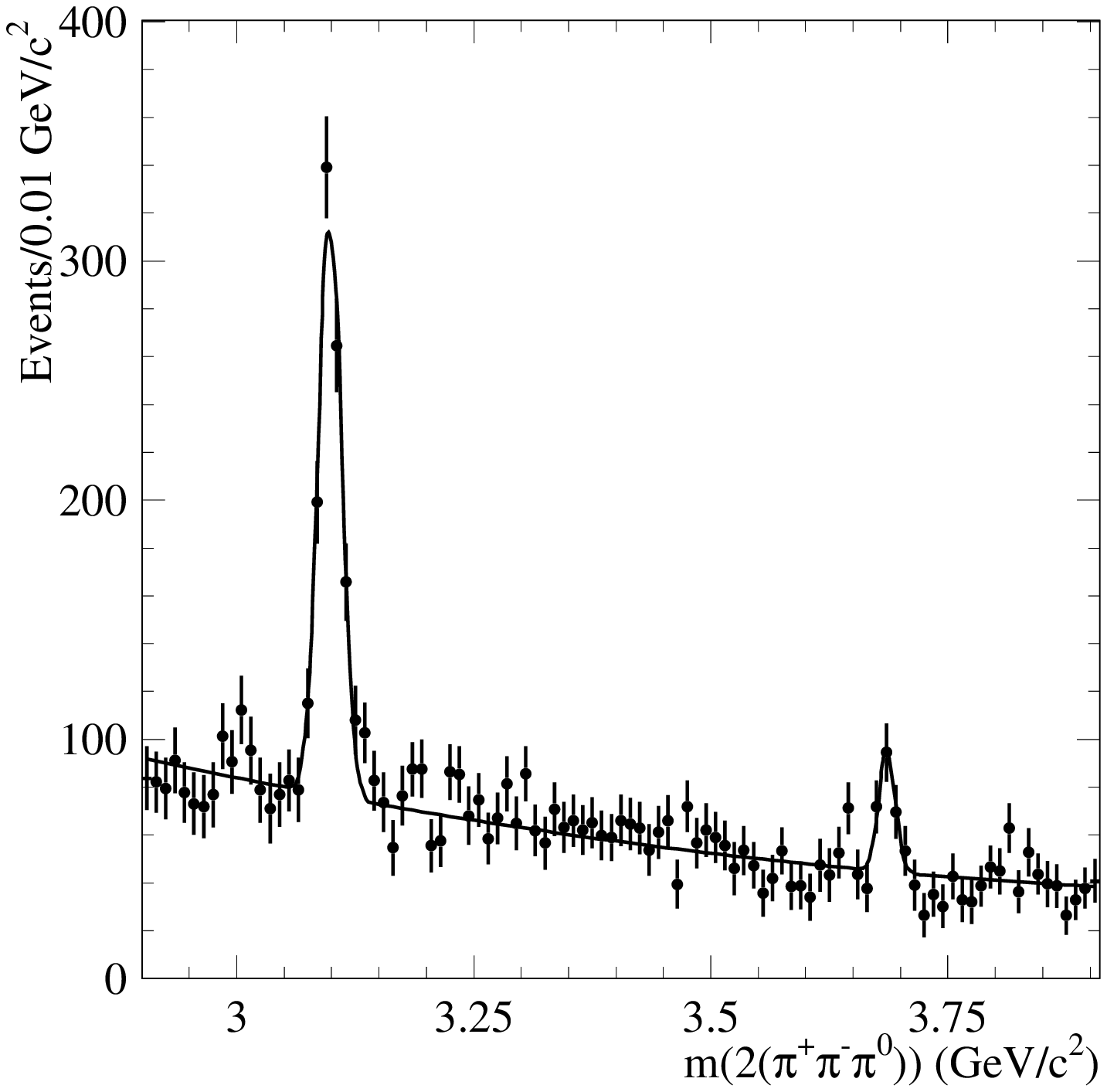}
\vspace{-0.4cm}
\caption{
The $2(\pipi\pi^0)$ mass 
distribution for ISR-produced events 
in the mass region around the $J/\psi$ and $\psi(2S)$;
there are clear signals at the $J/\psi$ and $\psi(2S)$ mass positions.
}
\label{jpsi_4pi2pi0}
\end{figure}
The observation of $J/\psi$ decaying into six-charged pions gives a direct
measurement of the 6$\pi$-mass resolution and the absolute energy scale. A
fit with a Gaussian for the $J/\psi$ peak and a polynomial function
for the continuum gives
$\sigma_{m(6\pi)}=8.7~\mevcc$ and less than 1~\mevcc difference
from the PDG~\cite{PDG} value for the $J/\psi$ mass. The observed mass resolution
agrees with the simulation within 10\%.

The observed $496\pm33$ events at the $J/\psi$ peak can be used to calculate
the branching fraction for $J/\psi\to 3(\pipi)$.  The simulation shows that
because of radiative effects only 90\% of the signal events are under the
Gaussian curve. Using the corrected number, we can calculate the products:
\begin{eqnarray*}
  B_{J/\psi\to 6\pi}\cdot\sigma_{\rm int}^{J/\psi}
  &=& \frac{N(J/\psi\to 3(\pipi))}%
           {d{\cal L}/dE\cdot\epsilon_{\rm MC}} \\
  &=& 57.1\pm3.8\pm3.4~\nb\mev\ ,\\
  B_{J/\psi\to 6\pi}\cdot\Gamma^{J/\psi}_{ee}
  &=& \frac{N(J/\psi\to 3(\pipi))\cdot m_{J/\psi}^2}%
           {6\pi^2\cdot d{\cal L}/dE\cdot\epsilon_{\rm MC}\cdot C} \\
  &=& (2.37\pm0.16\pm0.14)\times 10^{-2}~\kev\ ,\\
\end{eqnarray*}
where
$$\sigma_{\rm int}^{J/\psi} = 6\pi^2\Gamma^{J/\psi}_{ee}C/m_{J/\psi}^2 =
  12983\pm409 \mbox \nb\cdot\mev$$
is the integral over the $J/\psi$ excitation curve; $\Gamma^{J/\psi}_{ee}$ is
the electronic width; $d{\cal L}/dE = 65.6~\invnb/\mev$ is the ISR luminosity
at the $J/\psi$ mass; $\epsilon_{\rm MC} = 0.147$ is the detection
efficiency from simulation with the corrections and errors discussed in 
Sec.~\ref{sec:Systematics};
and  $C = 3.894\times 10^{11}~\nb\mev^2$ is a
conversion constant.  
The subscript ``$6\pi$'' for branching fractions refers
to the $3(\pipi)$ final state exclusively.

Using $\Gamma^{J/\psi}_{ee} =5.40\pm0.18~\kev$ ~\cite{PDG}, we obtain 
$B_{J/\psi\to 6\pi} = (4.40\pm 0.29\pm 0.29)\times 10^{-3}$, 
substantially more precise than the current PDG value
$B_{J/\psi\to 6\pi} = (4.0\pm 2.0)\times 10^{-3}$ ~\cite{PDG}. The systematic error
includes a 3\% uncertainty in $\Gamma^{J/\psi}_{ee}$.

The $\psi(2S)$ peak mostly corresponds to the decay chain
$\psi(2S)\to J/\psi\pipi\to 3(\pipi)$.
The number of events extracted from a fit to a Gaussian
distribution for the $\psi(2S)$ peak and
a polynomial function for the continuum is $61\pm 16$. The direct
decay of $\psi(2S)\to 3(\pipi)$ is very small and using $B_{\psi(2S)\to
6\pi} = (1.5\pm 1.0)\times 10^{-4}$ ~\cite{PDG} only 5 events are
expected -- less than the expected statistical fluctuation in the
observed number  of $\psi(2S)$ 
events. On the other hand using
$B_{\psi(2S)\to J/\psi\pipi} = 0.317\pm 0.011$ ~\cite{PDG}
and the recently measured  $B_{J/\psi\to 4\pi} = (3.61\pm 0.37)\times 
10^{-3}$~\cite{isr4pi}, 
we estimate that $42\pm 5$ events 
should be seen from the $\psi(2S)\to J/\psi\pipi\to 3(\pipi) $ decay
chain. This estimate may be  compared with a direct measurement of
$46\pm8$ events, using data  selected by taking events with four-pion
invariant mass combinations within $\pm 50$~\mevcc of the $J/\psi$
mass and counting the  number of $\psi(2S)$ events in the six-pion
mass distribution. Since the  total number of events from the six-pion
decay of the $\psi(2S)$   is less than 1$\sigma$ larger than the
number decaying through the  $J/\psi\pipi$ decay chain, no
significant measurement can  be made for the direct decay rate of
$\psi(2S)\to 3(\pipi)$ from these data.

Figure~\ref{jpsi_2k4pi} shows the $J/\psi$  and $\psi(2S)$ signals in the
$K^+ K^- 2(\pipi)$ mode. Background is not subtracted. The numbers of events under the 
Gaussian curves
are $205\pm 17$  and $51\pm 11$  respectively. 
As shown in Sec.~\ref{sec:2k4pi},   $35\pm 7$  $J/\psi$ events decay to the
$\phi 4\pi$ final state.
The mass resolution is
about 6.5~\mevcc for the $2K4\pi$ channel. Using $B_{\psi(2S)\to
J/\psi\pipi}$ and the recently measured $B_{J/\psi\to 2K2\pi} = (6.2\pm
0.7)\times 10^{-3}$~\cite{isr4pi} we can expect $25\pm 3$ events due
the decay chain $\psi(2S)\to J/\psi\pipi\to K^+ K^- 2(\pipi)$. 
The difference of $26\pm 13$ events between the total number of
$\psi(2S)\to K^+ K^- 2(\pipi)$  decays and this specific decay chain
provides an  estimate of direct
$\psi(2S)\to K^+ K^- 2(\pipi)$ decay.

Using the radiative correction factor 0.9, ISR luminosity $d{\cal
L}/dE = 84.0~\invnb/\mev$  at $\psi(2S)$ mass, and
$\epsilon_{2K4\pi}$  from
simulation with efficiency corrections, we obtain
\begin{eqnarray*}
\lefteqn{B_{J/\psi\to K^+K^- 2(\pipi)}\cdot\sigma_{\rm int}^{J/\psi}} \\
  &\qquad =& (66.2\pm5.5\pm4.0)~\nb\mev\ ,\\
\lefteqn{B_{J/\psi\to K^+K^- 2(\pipi)}\cdot\Gamma^{J/\psi}_{ee}} \\
  &\qquad =& (2.75\pm0.23\pm0.17)\times 10^{-2}~\kev\ ,\\
\lefteqn{B_{J/\psi\to \phi(1020) 2(\pipi)}\cdot B_{\phi\to
  K^+K^-}\cdot\Gamma^{J/\psi}_{ee}} \\ 
  &\qquad =& (0.47\pm0.09\pm0.04)\times 10^{-2}~\kev\ ,\\
\lefteqn{B_{\psi(2S)\to K^+K^- 2(\pipi)}\cdot\Gamma^{\psi(2S)}_{ee}} \\
  &\qquad =& (4.4\pm2.1\pm0.3)\times 10^{-3}~\kev\ .
\end{eqnarray*}
The systematic errors are mainly due to the uncertainties in the acceptance and
ISR luminosity. 

Using the world average values for $\Gamma^{J/\psi}_{ee}$, $\Gamma^{\psi(2S)}_{ee}$ and
$B_{\phi\to K^+K^-}$~\cite{PDG},  we calculate the branching fractions\\
\begin{eqnarray*}
  B_{J/\psi\to 2K4\pi} &=& (5.09\pm 0.42\pm 0.35)\times 10^{-3},\\
  B_{J/\psi\to\phi 4\pi} &=& (1.77\pm 0.35\pm 0.12)\times 10^{-3},\\
  B_{\psi(2S)\to 2K4\pi} &=& (2.1\pm 1.0\pm 0.2)\times 10^{-3},
\end{eqnarray*}
to be compared with the current world average values~\cite{PDG} of $(3.1\pm1.3)\times
10^{-3}$ for $J/\psi\to 2K4\pi$ and $(1.60\pm0.32)\times10^{-3}$ for
$J/\psi\to\phi 4\pi$. 
No earlier measurements are available for $\psi(2S)\to K^+ K^- 2(\pipi)$  decays.
The uncertainty in
$\Gamma^{J/\psi}_{ee}$ has been added in quadrature to the systematic error estimate.

The observation of $J/\psi$ decaying into four-charged and
two-neutral-pions gives a  direct
measurement of the $2(\pipi\pi^0)$-mass resolution and the absolute
energy scale in this mode. The mass distribution after background
subtraction is shown  in Fig.~\ref{jpsi_4pi2pi0}.  
A fit with a Gaussian for the $J/\psi$ peak and a polynomial function
for the continuum gives a mass resolution of
$\sigma_{m(2(\pipi\pi^0))}=12.7~\mevcc$.
The central value of the mass lies 1.5~\mevcc above the PDG~\cite{PDG}
value for the $J/\psi$ mass and gives an estimate of systematic
uncertainty in absolute energy scale determined by invariant mass of
six pions. The observed mass resolution
agrees with the simulation very well.

The $761\pm42$ events observed at the $J/\psi$ peak 
 can be used to calculate
the branching fraction for $J/\psi\to 2(\pipi\pi^0)$ as was done
 for  $J/\psi\to 3(\pipi)$.
  We calculate the products:
\begin{eqnarray*}
  B_{J/\psi\to 4\pi 2\pi^0}\cdot\sigma_{\rm int}^{J/\psi}
  &=& \frac{N(J/\psi\to 4\pi 2\pi^0)}%
           {d{\cal L}/dE\cdot\epsilon_{\rm MC}} \\
  &=& 215\pm12\pm24~\nb\mev\ ,\\
  B_{J/\psi\to 4\pi 2\pi^0}\cdot\Gamma^{J/\psi}_{ee}
  &=& \frac{N(J/\psi\to 4\pi 2\pi^0)\cdot m_{J/\psi}^2}%
           {6\pi^2\cdot d{\cal L}/dE\cdot\epsilon_{\rm MC}\cdot C} \\
  &=& (8.9\pm0.5\pm1.0)\times 10^{-2}~\kev\ ,\\
\end{eqnarray*}
where  $\epsilon_{\rm MC} = 0.060$ is the detection
efficiency from simulation with the corrections and error discussed in
Sec.~\ref{sec:Systematics2}. 
The subscript ``$4\pi 2\pi^0$'' for branching fractions refers
to the $2(\pipi\pi^0)$ final state exclusively.

Using $\Gamma^{J/\psi}_{ee} =5.40\pm0.18~\kev$ ~\cite{PDG}, we obtain the
result $B_{J/\psi\to 4\pi 2\pi^0} = (1.65\pm 0.10\pm
0.18)\times10^{-2}$.  No entry for this branching fraction exists 
in  PDG~\cite{PDG}. 
As was noted in Sec.~\ref{Physics2}  $170\pm24$
events are observed in
the $J/\psi\to\omega\pipi\pi^0$ channel. This corresponds to 
the value,
\begin{eqnarray*}
B_{J/\psi\to\omega\pipi\pi^0}\cdot\Gamma^{J/\psi}_{ee}
 &=& (2.2\pm0.3\pm0.2)\times 10^{-2}~\kev\ ,
\end{eqnarray*}
including the decay rate for $\omega\to\pipi\pi^0$. The
corresponding fraction is $B_{J/\psi\to\omega\pipi\pi^0}
= (0.41\pm 0.06\pm0.04)\times10^{-2}$, 
which  is the first measurement of this branching
fraction. There are  also $13\pm3.6$ events from the  
$J/\psi\to\omega\eta$ decay. Taking into account the decay rates
of $\omega$ and $\eta$ to $\pipi\pi^0$, we obtain
$B_{J/\psi\to\omega\eta} = (1.47\pm 0.41\pm 0.15)\times 10^{-3}$, 
in agreement with the current PDG value $B_{J/\psi\to\omega\eta} =
(1.58\pm 0.16)\times 10^{-3}$.  

The $\psi(2S)$ peak can partly correspond to the decay chains
$\psi(2S)\to J/\psi\ppz\to 2(\pipi\pi^0)$ or $\psi(2S)\to J/\psi\pipi\to 
2(\pipi\pi^0)$.
Using $B_{J/\psi\to 2(\pipi)} = (3.61\pm 0.37)\times10^{-3}$ from 
Ref.~\cite{isr4pi}, and assuming that $B_{J/\psi\to \pipi\ppz}$ is at the same 
level, as confirmed by a preliminary study of that channel,
we estimate that $43\pm15$ events from the above decay chains contribute 
to the $\psi(2S)$ peak.
The total number of $\psi(2S)$  signal events extracted from a fit to a Gaussian
distribution for the $\psi(2S)$ peak and
a polynomial function for the continuum is $128\pm 20$.
The $43\pm15$ events estimated above should be subtracted from this
total to  obtain an estimate for the number of direct
$\psi(2S)\to 2(\pipi\pi^0)$ decays.

Using  $d{\cal L}/dE = 84.0~\invnb/\mev$ as the ISR luminosity
at the $\psi(2S)$ mass; $\epsilon_{\rm MC} = 0.059$ as the detection
efficiency from simulation with corrections and systematic error discussed in
Sec.~\ref{sec:Systematics2}, and a 0.90 correction factor we obtain
\begin{eqnarray*}
B_{\psi(2S)\to 4\pi 2\pi^0}\cdot\sigma_{\rm int}^{\psi(2S)}
&=& \frac{N(\psi(2S)\to 4\pi 2\pi^0)}%
           {d{\cal L}/dE\cdot\epsilon_{\rm MC}} \\
&=& 19.1\pm5.6\pm2.2~\nb\mev\ ,\\
B_{\psi(2S)\to 4\pi 2\pi^0}\cdot\Gamma^{\psi(2S)}_{ee}
&=& \frac{N(\psi(2S)\to 4\pi 2\pi^0)\cdot m_{\psi(2S)}^2}%
           {6\pi^2\cdot d{\cal L}/dE\cdot\epsilon_{\rm MC}\cdot C} \\
&=&(1.12\pm0.33\pm0.13)\times 10^{-2}~\kev.\\
\end{eqnarray*}
 Using the value of $\Gamma^{\psi(2S)}_{ee}$ from the PDG~\cite{PDG} we 
calculate 
$B_{\psi(2S)\to 4\pi 2\pi^0} = (5.3 \pm 1.6 \pm 0.6) \times
10^{-3}$. 
This is the first measurement of that branching ratio.

\section{\boldmath The $J/\psi$ production rate and continuum}
Having measured products of $\sigma_{\rm int}^{J/\psi}$ and 
branching
fractions for $J/\psi$  decaying to $6\pi$ and $2K4\pi$, it is interesting
to compare them with the non-resonant cross sections 
(continuum) at that energy.
Using a second order polynomial approximation of the cross sections from
Tables~\ref{6pi_tab}, \ref{2k4pi_tab}, and \ref{4pi2pi0_tab} around the $J/\psi$
peak within $\pm 0.4\gev$ (events from the peak are excluded), 
the following cross sections are obtained for the continuum at the
$J/\psi$ mass:
\begin{eqnarray*}
  \sigma_{6\pi}   &=& 0.54\pm0.03~\nb \\
  \sigma_{4\pi 2\pi^0}   &=& 2.17\pm0.04~\nb \\
  \sigma_{2K4\pi} &=& 0.54\pm0.04 \nb\ .
\end{eqnarray*}
Table~\ref{jpsi_to_xs} presents the
ratios $B_{J/\psi\to f}\cdot\sigma_{\rm int}^{J/\psi}/\sigma_{\epem\to f}$ for $f=6\pi,
2K4\pi$.  In these ratios most of the experimental systematic errors cancel. 
Also shown are the ratios for the $J/\psi\to\mumu$ final
state~\cite{Druzhinin1}, and $J/\psi\to 2(\pipi)$, 
$J/\psi\to K^+ K^-\pipi$, and $J/\psi\to K^+ K^- K^+ K^-$~\cite{isr4pi}.
\begin{table}[tbh]
\caption{
Ratios of the $J/\psi$ partial production rates to continuum cross
sections. The result for $\mumu$ is from Ref.~\cite{Druzhinin1}.
The results for $2(\pipi)$, $K^+ K^- \pipi$ and $K^+ K^- K^+ K^-$ are
from  Ref.~\cite{isr4pi}.
}
\label{jpsi_to_xs}
\begin{ruledtabular}
\begin{tabular}{cc}
Final state, $f$ & 
$B_{J/\psi\to f}\cdot\sigma_{\rm int}^{J/\psi} / \sigma_{\epem\to f}$
(MeV) \\ \hline
$3(\pipi)$ & $106 \pm 10$ \\
$2(\pipi\pi^0)$ & $99.1 \pm 6.5$ \\
$K^+ K^- 2(\pipi)$ & $122 \pm 10$ \\
$2(\pipi)$ & $85.1 \pm 7.9$ \\
$K^+ K^- \pipi$ & $166 \pm 19$ \\ 
$K^+ K^- K^+ K^-$ & $138 \pm 32$ \\
$\mumu$ & $84.12 \pm 0.67$ 
\end{tabular}
\end{ruledtabular}
\end{table}

The strong decay of the $J/\psi$ to an even number of pions is
forbidden by G-parity conservation
and therefore this decay is expected to be dominated by a single photon.
No such suppression due to G-parity occurs for the strong decay of the $J/\psi$
for the other modes.
The ratio obtained for both of the $6\pi$ final states is 
close to
that for $\mumu$ and for $4\pi$, indicating that the single-photon 
exchange dominates 
for the $J/\psi$ decays into these modes.
For the $J/\psi$ decays to final states with kaons, the single-photon
mechanism may  be less dominant, as indicated by the significantly
larger values  of the ratios.

\section{Summary}
\label{sec:Summary}
\noindent
The large 232~\invfb data sample accumulated near the $\Upsilon(4S)$,
together with  the excellent resolution, charged particle
identification, and open  trigger of the \babar\ detector, affords a
unique opportunity  to study the $3(\pipi)$, $2(\pipi\pi^0)$, and
$K^+K^- 2(\pipi)$  final states produced at low effective $\epem$
c.m.\@ energy via  ISR.  Not only are the data samples extraordinarily
large and  well measured, but they do not suffer from the relative
normalization  uncertainties which seem to plague direct measurements
from earlier storage rings such as DCI and SPEAR.

Since the luminosity and efficiencies are understood within 3-5\%
accuracy, these data  allow useful measurements of $R$, the ratio of the
hadronic to  di-muon cross sections, in the low $\epem$ energy regime,
providing  important input for calculating corrections needed for muon
$g-2$  measurements.  Cross section measurements for the reactions
$\epem\to 3(\pipi)$ and $\epem\to 2(\pipi\pi^0)$, with about 6\% and 10\% total
systematic errors,  respectively, are significant improvements over
earlier  measurements, while no earlier measurements exist for the
reaction $\epem\to K^+K^- 2(\pipi)$.

These final states also provide new information on hadron
spectroscopy. Structure is  observed around 1.9~\gevcc, in both the
$3(\pipi)$  and $2(\pipi\pi^0)$ final states, which is similar to that
previously  observed by the DM2 and FOCUS experiments. A simple
coherent model fits  the structure well, and leads to the following
parameters: $m = 1.87 \pm 0.02~\gevcc$, $\Gamma = 0.14 \pm
0.03~\gev$ with $-3 \pm 15^o $  relative phase to the
continuum. However, these values are model dependent.

Resonance substructure is also seen in the intermediate mass
combinations for these  final states. The $\pipi$ decay of the $\rho(770)$
dominates  the substructure observed in the $3(\pipi)$ final state,
with little  evidence for any other structures. A model with a single
$\rho(770)$  describes the observed distributions very well. On the
other hand, not  only does the $2(\pipi\pi^0)$ final state display
contributions from $\rho(770) 4\pi$  but also from $\omega(782)\pipi\pi^0$ 
and $\omega(782)\eta$ final states. Structures that may correspond to
$f_0 (980)$, $f_0 (1370)$ and/or $f_2 (1270)$
are also visible in intermediate mass  combinations. A detailed
understanding of  this complex structure will probably require a full
PWA using all of the six-pion final states.

Except for a structure around 1.6~\gev, the ratio of 
$\epem\to 2(\pipi\pi^0)$ to $\ep\en\to 3(\pipi)$
cross  sections is equal to $3.98\pm0.06\pm0.41$ over a wide range of
c.m.\@ energies.  However, it is no longer flat and substantially
smaller when  the contribution from $\omega(782)$ decays to
$\pipi\pi^0$ is excluded. These results can indicate
and help to understand more complex structures existing in the six-pion
final states. 

A resonance-like structure is seen in the $\ep\en\to\omega(782)\eta$
cross section  around 1.6~\gev. A resonance fit gives 
 $m = 1.645\pm0.008~\gevcc$, $\Gamma =
0.114\pm0.014~\gev$ and peak cross section $\sigma_0 = 3.08 \pm 0.33 $~nb. 
This object might be associated with the 
$\omega(1650)$, for which a $\omega(782)\eta$ final
state has been seen~\cite{PDG}, 
but the width measured here is substantially smaller. The observed
parameter values  are actually much closer to those of the 
$\phi(1680)$~\cite{PDG}, but there are no observations for an
$\omega(782)\eta$ decay mode of the $\phi(1680)$.

The energy dependence of the cross section for the reaction $\epem\to K^+
K^- 2(\pipi)$ from threshold to 4.5~\gev is measured 
with about 15\% systematic uncertainty.
There is clear evidence for resonance sub-structure including the 
$K^{*0}(892)$ and $\phi$.

These data also allow a study of $J/\psi$ and $\psi(2S)$ production,
and the measurement of the product of decay branching fractions and
the \epem width of the $J/\psi$ with the best accuracy to date. The results
are as follows:
\begin{eqnarray*}
\lefteqn{B_{J/\psi\to 3(\pipi)}\cdot\Gamma^{J/\psi}_{ee}} \\
  &\qquad =& (2.37\pm0.16\pm0.14)\times 10^{-2}~\kev\ ,\\
\lefteqn{B_{J/\psi\to K^+K^- 2(\pipi)}\cdot\Gamma^{J/\psi}_{ee}} \\
  &\qquad =& (2.75\pm0.23\pm0.17)\times 10^{-2}~\kev\ ,\\
\lefteqn{B_{J/\psi\to \phi(1020) 2(\pipi)}\cdot B_{\phi\to
K^+K^-}\cdot\Gamma^{J/\psi}_{ee}}  \\
  &\qquad =& (0.47\pm0.09\pm0.03)\times 10^{-2}~\kev\ ,\\
\lefteqn{B_{J/\psi\to 2(\pipi\pi^0)}\cdot\Gamma^{J/\psi}_{ee}} \\
  &\qquad =& (8.9\pm0.5\pm1.0)\times 10^{-2}~\kev\ ,\\
\lefteqn{B_{J/\psi\to\omega\pipi\pi^0}\cdot\Gamma^{J/\psi}_{ee}} \\
  &\qquad =& (2.2\pm0.3\pm0.2)\times 10^{-2}~\kev\ ,\\
\lefteqn{B_{\psi(2S)\to 2(\pipi\pi^0)}\cdot\Gamma^{\psi(2S)}_{ee}} \\
  &\qquad =& (1.12\pm0.33\pm0.12)\times 10^{-2}~\kev\ ,\\
\lefteqn{B_{\psi(2S)\to K^+K^- 2(\pipi)}\cdot\Gamma^{\psi(2S)}_{ee}} \\
  &\qquad =& (4.4\pm2.1\pm0.3)\times 10^{-3}~\kev\ .
\end{eqnarray*}

Dominance of the single-photon decay-mechanism for
$J/\psi\to 3(\pipi)$ and $J/\psi\to 2(\pipi\pi^0)$ decays has been
demonstrated by comparison  with the continuum
cross sections for $\epem\to 3(\pipi)$ and  $\epem\to 2(\pipi\pi^0)$.

\section{Acknowledgments}
\label{sec:Acknowledgments}

We are grateful for the 
extraordinary contributions of our \pep2\ colleagues in
achieving the excellent luminosity and machine conditions
that have made this work possible.
The success of this project also relies critically on the 
expertise and dedication of the computing organizations that 
support \babar.
The collaborating institutions wish to thank 
SLAC for its support and the kind hospitality extended to them. 
This work is supported by the
US Department of Energy
and National Science Foundation, the
Natural Sciences and Engineering Research Council (Canada),
Institute of High Energy Physics (China), the
Commissariat \`a l'Energie Atomique and
Institut National de Physique Nucl\'eaire et de Physique des Particules
(France), the
Bundesministerium f\"ur Bildung und Forschung and
Deutsche Forschungsgemeinschaft
(Germany), the
Istituto Nazionale di Fisica Nucleare (Italy),
the Foundation for Fundamental Research on Matter (The Netherlands),
the Research Council of Norway, the
Ministry of Science and Technology of the Russian Federation, and the
Particle Physics and Astronomy Research Council (United Kingdom). 
Individuals have received support from 
CONACyT (Mexico), the Marie-Curie Intra European Fellowship program (European Union),
the A. P. Sloan Foundation, 
the Research Corporation,
and the Alexander von Humboldt Foundation.

\end{document}